\documentclass[acmsmall,screen]{acmart}
\usepackage{pifont}
\usepackage{tikz}
\usepackage{mdframed}

\usepackage[normalem]{ulem}

\usepackage{xcolor}

\usepackage[commandnameprefix=always]{changes}  
\definechangesauthor[name={yulei}, color=red]{author1}

\AtBeginDocument{%
  }

\setcopyright{cc}
\setcctype{by}
\acmDOI{10.1145/3728878}
\acmYear{2025}
\acmJournal{PACMSE}
\acmVolume{2}
\acmNumber{ISSTA}
\acmArticle{ISSTA009}
\acmMonth{7}
\received{2024-10-31}
\received[accepted]{2025-03-31}

\begin{document}

\title[Reinforced Large Language Model for Explainable Smart Contract Vulnerability Detection]{Smart-LLaMA-DPO: Reinforced Large Language Model for Explainable Smart Contract Vulnerability Detection}

\author{Lei Yu}
\authornote{Affiliated with University of Chinese Academy of Sciences, Beijing, China.}
\authornote{Affiliated with Integrative Innovation Center, Institute of Software, Chinese Academy of Sciences, Beijing, China.}
\authornote{Affiliated with Key Laboratory of System Software (Chinese Academy of Sciences) and State Key Laboratory of Computer Science, Institute of Software, Chinese Academy of Sciences, Beijing, China.}

\orcid{0000-0003-3134-3746}
\affiliation{
  \institution{Institute of Software, Chinese Academy of Sciences}
  \city{Beijing}
  \country{China}
}
\email{yulei2022@iscas.ac.cn}

\author{Zhirong Huang}
\authornotemark[1]
\authornotemark[2]
\authornotemark[3]
\orcid{0009-0007-9652-9097}
\affiliation{
  \institution{Institute of Software, Chinese Academy of Sciences}
  \city{Beijing}
  \country{China}
}
\email{huangzhirong2022@iscas.ac.cn}

\author{Hang Yuan}
\authornotemark[1]
\authornote{Affiliated with Laboratory of Precise Computing, Institute of Software, Chinese Academy of Sciences, Beijing, China.}
\orcid{0009-0001-9607-5538}
\affiliation{
  \institution{Institute of Software, Chinese Academy of Sciences}
  \city{Beijing}
  \country{China}
}
\email{yuanhang2023@iscas.ac.cn}

\author{Shiqi Cheng}
\authornotemark[4]
\orcid{0009-0008-1407-8446}
\affiliation{
  \institution{Institute of Software, Chinese Academy of Sciences}
  \city{Beijing}
  \country{China}
}
\email{chengshiqi@iscas.ac.cn}

\author{Li Yang}
\authornotemark[4]
\authornote{Li Yang and Fengjun Zhang are the corresponding authors.}
\orcid{0000-0001-8364-6525}
\affiliation{
  \institution{Institute of Software, Chinese Academy of Sciences}
  \city{Beijing}
  \country{China}
}
\email{yangli2017@iscas.ac.cn}

\author{Fengjun Zhang}
\authornotemark[2]
\authornotemark[3]
\authornotemark[5]
\orcid{0000-0002-3830-8786}
\affiliation{
  \institution{Institute of Software, Chinese Academy of Sciences}
  \city{Beijing}
  \country{China}
}
\email{fengjun@iscas.ac.cn}

\author{Chenjie Shen}
\authornotemark[1]
\authornotemark[4]
\orcid{0009-0000-6212-9721}
\affiliation{
  \institution{Institute of Software, Chinese Academy of Sciences}
  \city{Beijing}
  \country{China}
}
\email{shenchenjie22@mails.ucas.ac.cn}

\author{Jiajia Ma}
\authornotemark[2]
\orcid{0000-0002-6028-4186}
\affiliation{
  \institution{Institute of Software, Chinese Academy of Sciences}
  \city{Beijing}
  \country{China}
}
\email{majiajia@iscas.ac.cn}

\author{Jingyuan Zhang}
\authornotemark[1]
\authornotemark[2]
\authornotemark[3]
\orcid{0000-0001-5475-3815}
\affiliation{
  \institution{Institute of Software, Chinese Academy of Sciences}
  \city{Beijing}
  \country{China}
}
\email{zhangjingyuan2023@iscas.ac.cn}

\author{Junyi Lu}
\authornotemark[1]
\authornotemark[4]
\orcid{0009-0006-5453-0316}
\affiliation{
  \institution{Institute of Software, Chinese Academy of Sciences}
  \city{Beijing}
  \country{China}
}
\email{lujunyi2022@iscas.ac.cn}

\author{Chun Zuo}
\orcid{0000-0002-0753-7164}
\affiliation{
  \institution{Sinosoft Company Limited}
  \city{Beijing}
  \country{China}
}
\email{zuochun@sinosoft.com.cn}

\renewcommand{\shortauthors}{Yu et al.}

\makeatletter
\def\@makefnmark{}
\makeatother

\begin{abstract}

Smart contract vulnerability detection is a critical challenge in the rapidly evolving blockchain landscape. Existing vulnerability detection methods face two main issues: (1) Existing datasets lack comprehensiveness and sufficient quality, with limited vulnerability type coverage and insufficient distinction between high-quality and low-quality explanations for preference learning. (2) Large language models (LLMs) often struggle with accurately interpreting specific concepts in smart contract security. Through our empirical analysis, we found that even after continual pre-training and supervised fine-tuning, LLMs still exhibit limitations in precisely understanding the execution order of state changes in smart contracts, which can lead to incorrect vulnerability explanations despite making correct detection decisions. These limitations result in poor detection performance, leading to potentially severe financial losses. To address these challenges, we propose Smart-LLaMA-DPO, an advanced detection method based on the LLaMA-3.1-8B. First, we construct a comprehensive dataset covering four vulnerability types and machine-unauditable vulnerabilities, containing labels, detailed explanations, and precise vulnerability locations for Supervised Fine-Tuning (SFT), as well as paired high-quality and low-quality outputs for Direct Preference Optimization (DPO). Second, we perform continual pre-training using large-scale smart contract code to enhance the LLM's understanding of specific security practices in smart contracts. Futhermore, we conduct supervised fine-tuning with our comprehensive dataset. Finally, we apply DPO, which leverages human feedback to improve the quality of generated explanations. Smart-LLaMA-DPO utilizes a specially designed loss function that encourages the LLM to increase the probability of preferred outputs while decreasing the probability of non-preferred outputs, thereby enhancing the LLM's ability to generate high-quality explanations. We evaluate Smart-LLaMA-DPO on four major vulnerability types: reentrancy, timestamp dependence, integer overflow/underflow, and delegatecall, as well as machine-unauditable vulnerabilities. Our method significantly outperforms state-of-the-art baselines, with average improvements of 10.43\% in F1 score and 7.87\% in accuracy. Moreover, both LLM evaluation and human evaluation demonstrate the superior quality of explanations generated by Smart-LLaMA-DPO in terms of correctness, thoroughness, and clarity.

\end{abstract}


\maketitle

\section{Introduction}
Blockchain technology has been rapidly adopted across various domains due to its decentralized architecture \cite{swan2015blockchain}. This innovative technology enables the creation of secure, distributed digital ledgers for recording transactions \cite{hewa2021survey}. By utilizing advanced cryptographic methods, blockchain ensures the integrity and verification of each transaction\cite{wood2014ethereum, yu2023money}. Within this ecosystem, smart contracts function as self-executing programs on the blockchain, automating the management of digital assets such as cryptocurrencies. These contracts are activated when specific conditions are met and, once deployed, become permanent components of the blockchain \cite{zou2019smart}. However, the immutability and inherent complexity of smart contracts pose significant security challenges \cite{zou2019smart}. The well-known DAO incident \cite{dhillon2017dao,mehar2019understanding} serves as a cautionary example, demonstrating the potential severity of such vulnerabilities. This security breach resulted in the illegal transfer of \$60 million worth of Ethereum, causing widespread impact on the blockchain community \cite{alharby2017blockchain, hegedHus2018towards}.

Researchers have developed various techniques for smart contract vulnerability detection, such as symbolic execution (Oyente \cite{luu2016making}, Mythril \cite{mueller2017mythril}) and static analysis (Slither \cite{feist2019slither}, SmartCheck \cite{tikhomirov2018smartcheck}). However, these methods often rely on predefined patterns and perform poorly in complex scenarios. Qian et al. \cite{qian2023cross} proposed cross-modality learning approaches that leverage information from both bytecode and source code. Recently, researchers have begun to explore the potential of LLMs in smart contract vulnerability detection and explanation (iAudit \cite{ma2024combining}). While iAudit shows promise in detecting logic vulnerabilities, its two-stage architecture that separates detection and explanation tasks can lead to inconsistencies. For instance shown in Fig. \ref{smart_case}, the detector incorrectly identifies a reentrancy vulnerability in the contract and misinterprets the sequence of external calls and state changes. While the Price Oracle Manipulation vulnerability is correctly identified, it is incorrectly attributed to insufficient access control rather than the critical issue of reliance on the Uniswap pool's instant price. Such inconsistencies, although mitigated by iAudit's Ranker-Critic mechanism, could still affect developers' remediation decisions. We conducted a systematic evaluation of existing smart contract vulnerability datasets, as shown in Table \ref{dataset_comparison}. While most existing datasets (A-D) provide only basic vulnerability labels without detailed explanations or location information, iAudit made progress by including these aspects. However, iAudit's dataset primarily focuses on logic vulnerabilities, and its data enhancement process relies on LLMs without human expert verification after LLM generation. Furthermore, the current datasets lack the structured format to support Direct Preference Optimization (DPO) training \cite{rafailov2023direct}, which significantly affects the LLM's ability to generate high-quality vulnerability explanations that match human expert standards.

Beyond iAudit, general LLMs also face significant challenges in the domain of smart contract vulnerability detection. They often encounter difficulties in dealing with smart contract-specific concepts and security implications. As shown in Fig. \ref{tab:motivation}, when faced with two explanations, one being an incorrect explanation from a general LLM (LLaMA3.1-8B-Instruct) and the other a partially correct explanation from Smart-LLaMA-DPO (only after continual pre-training and supervised fine-tuning), the general LLM incorrectly identifies a reentrancy vulnerability in the contract. It misinterprets the meaning of external calls in the 'buyInternal()' function, failing to recognize that reentrancy vulnerabilities typically occur when contract state or balance changes are made after external calls, which is not the case in this contract. Smart-LLaMA-DPO (only after continual pre-training and supervised fine-tuning) correctly identifies the contract's security but still misunderstands the order of operations in its explanation. This subtle but critical misunderstanding highlights the limitations of relying solely on continual pre-training and supervised fine-tuning.

To address these challenges, we propose Smart-LLaMA-DPO, based on the LLaMA-3.1-8B model. This approach combines continual pre-training, supervised fine-tuning, and direct preference optimization (DPO). Unlike iAudit’s two-stage architecture, which may lead to inconsistencies between detection and explanation, our approach leverages Direct Preference Optimization (DPO) \cite{rafailov2023direct} to align these tasks within a unified framework, enabling more consistent and context-aware results. By ultilizing large-scale smart contract code for domain-specific pre-training, we enable the LLM to better understand the syntax and semantics of smart contracts, overcoming the limitations of general LLMs in understanding contract semantics as shown in Fig. \ref{tab:motivation}. Subsequently, we construct a comprehensive smart contract vulnerability dataset containing detailed explanations and precise location information. The fine-tuning process utilizing this high-quality dataset enhances the LLM's capabilities, enabling it to not only detect vulnerabilities but also generate explanations. This approach effectively addresses the limitations of existing datasets as illustrated in Table \ref{dataset_comparison}. Finally, to address the issues in explanation quality that persist even after CPT and SFT training (as seen in Smart-LLaMA-DPO (only after CPT and SFT)'s partially incorrect explanation in Fig. \ref{tab:motivation}), we innovatively introduce DPO \cite{rafailov2023direct}. By constructing a dataset containing pairs of model outputs with varying quality, DPO enables the LLM to learn to generate higher quality explanations that better align with human expert expectations. Each pair in this dataset consists of two outputs: one preferred by human experts, representing high-quality explanations, and another of lower quality. The core of the Smart-LLaMA-DPO approach lies in learning directly from expert preferences, without the need for explicit reward modeling or complex reinforcement learning processes. Smart-LLaMA-DPO uses a specially designed loss function that encourages the LLM to increase the probability of preferred outputs while decreasing the probability of non-preferred outputs.

In the data construction process, we used Qwen2.5-72B-Instruct and Mistral-Large-Instruct-2407-123B to generate initial explanations, scored by Llama-3.1-70B-Instruct. Smart contract security experts refined high-scoring explanations to create the SFT dataset. For the DPO dataset, experts rewrote lower-scored outputs into "suboptimal" versions, which, while correct, lack depth and clarity compared to high-quality outputs, creating a reasonable quality gap.

To validate the effectiveness of Smart-LLaMA-DPO, we conducted a comprehensive experimental evaluation. We evaluated the Smart-LLaMA-DPO framework on a challenging dataset \cite{qian2023cross} covering four major vulnerability types (reentrancy, timestamp dependence, integer overflow/underflow, and delegatecall) and machine unauditable vulnerabilities (seven types) from \cite{zhang2023critical}. The results show that Smart-LLaMA-DPO significantly outperforms state-of-the-art methods across all vulnerability types. Smart-LLaMA-DPO achieves F1 scores 7.51\%, 1.54\%, 11.07\%, and 6.06\% higher than the previous best performance for reentrancy, timestamp dependence, integer overflow/underflow, and delegatecall vulnerabilities, respectively. In terms of accuracy, Smart-LLaMA-DPO surpasses the previous SOTA methods by 5.65\%, 1.02\%, 10.52\%, and 3.90\% for these four vulnerability types. For machine unauditable vulnerabilities, Smart-LLaMA-DPO surpasses the previous best baseline (iAudit) with an increase of 25.98\% in F1 score and 18.25\% in accuracy. Both LLM evaluation and human evaluation show that Smart-LLaMA-DPO produces more accurate, comprehensive, and concise explanations than baselines. Human evaluation gave positive scores (4 or 3 points) in 81.15\%, 83.88\%, and 94.63\% for correctness, thoroughness, and clarity.

The main contributions of this paper are as follows:
\begin{itemize}
\item To the best of our knowledge, we are the first to introduce preference-based optimization in smart contract vulnerability detection and explanation.
\item We propose Smart-LLaMA-DPO, a new method combining continual pre-training, supervised fine-tuning and direct preference optimization for smart contract vulnerability detection, achieving state-of-the-art performance across four major vulnerability types and machine unauditable vulnerabilities (seven types).
\item We validate the effectiveness of Smart-LLaMA-DPO in generating high-quality explanations through both LLM evaluation and human evaluation.
\end{itemize}

\section{Background and Motivation}

\subsection{Key Terms}

\textbf{Solidity \cite{solidity_docs}} is the primary programming language for Ethereum smart contracts. \textbf{Call.value()} is a low-level function for sending Ether that may introduce reentrancy vulnerabilities \cite{solidity_call}. \textbf{Delegatecall} executes target contract in the caller's context, useful for upgradeable contracts but potentially dangerous \cite{solidity_delegatecall}. \textbf{Block.timestamp} is the block's timestamp that may be manipulated by miners \cite{solidity_timestamp}. \textbf{SmartBugs dataset \cite{ferreira2020smartbugs}} is a curated dataset of vulnerable Ethereum smart contracts used for security research. \textbf{Qwen2.5-72B-Instruct \cite{qwen2.5}} developed by Alibaba and \textbf{Mistral-Large-Instruct-2407-123B \cite{mistrallarge2}} developed by Mistral AI are two powerful instruction-tuned LLMs.

\subsection{Problem Statement}

We propose an automated approach to detect vulnerabilities in smart contracts and provide explanations. Our method assigns a label $\hat{y}$ to each independent smart contract, where $\hat{y}=1$ indicates the presence of a vulnerability and $\hat{y}=0$ denotes security. We adopts the smart contract vulnerability taxonomy widely used in several recent studies \cite{yu2024smart, qian2023cross, zhang2023demystifying}, focusing on four key vulnerability types and seven machine-unauditable vulnerabilities.

\textbf{Reentrancy vulnerability} (RE) occurs when a contract calls an external contract or transfers assets (Ether or tokens) before completing internal state changes, allowing attackers to repeatedly call the vulnerable function and potentially withdraw funds multiple times.

\textbf{Timestamp dependence vulnerability} (TD) occurs when smart contracts rely on block timestamps for critical operations. Miners can manipulate these timestamps, potentially compromising contract integrity and leading to financial losses. This vulnerability often affects contracts using timestamps for random number generation or key decision-making processes.

\textbf{Integer Overflow/Underflow} (IO) occurs when the result of an arithmetic operation exceeds the storage range of the variable. In an overflow, the value "wraps around" to the minimum value for that type, while in an underflow, it "wraps around" to the maximum value. This can lead to unexpected contract behavior such as incorrect balances or out-of-control loops.

\textbf{Delegatecall} (DE) is a low-level function call that allows a contract to dynamically load code from another contract. While this provides powerful upgradeability, it can lead to severe security vulnerabilities if used improperly. The main risk is that the called contract executes in the context of the calling contract and can thus modify the calling contract's storage.

\textbf{Machine-unauditable Vulnerabilities} (MU) represent vulnerabilities that are difficult to detect through automated tools and require domain expertise to identify. These include: \textbf{Price Oracle Manipulation} (PO) from improper use of price oracle APIs, \textbf{Erroneous Accounting} (EA) from incorrect business model calculations, \textbf{ID Uniqueness Violations} (IU) from failures in ensuring unique identifiers, \textbf{Inconsistent State Updates} (IS) from incorrect correlated state variable updates, \textbf{Privilege Escalation} (PE) from insufficient access control, \textbf{Atomicity Violations} (AV) from interfering concurrent flows, and \textbf{Contract Implementation Specific} (CI) vulnerabilities requiring implementation context understanding.

\textbf{We focus on these four vulnerability types and machine-unauditable vulnerabilities} for the following reasons: (i) These four vulnerabilities account for approximately 70\% of financial losses and occur with higher frequency in Ethereum smart contracts \cite{chen2020survey,gao2019easyflow,praitheeshan2019security}. (ii) According to OWASP 2023 Smart Contract Top 10~\cite{owasp2023}, three of them are ranked as the top-3 vulnerabilities, with delegatecall also included as unchecked external calls. (iii) Recent research \cite{zhang2023demystifying} shows machine-unauditable vulnerabilities represent a significant security challenge, with price oracle manipulation causing at least \$44.8M in losses (34\% of real-world exploits).

\subsection{Motivations}

In this section, we analyze the motivations for improving smart contract vulnerability detection. Using real-world examples and datasets, we illustrate the limitations of current methods, emphasizing the need for high-quality datasets and DPO training.

\begin{table}[h]
\caption{Comparison of Smart Contract Vulnerability Datasets.}
\label{dataset_comparison}
\centering
\begin{tabular}{|l|c|c|c|c|c|}
\hline
\textbf{Dataset} & \textbf{Label} & \textbf{Explanation} & \textbf{Location} & \textbf{DPO Format} & \textbf{Types} \\
\hline
Dataset A \cite{wu2021peculiar} & $\checkmark$ & \ding{55} & \ding{55} & \ding{55} & 1 \\
\hline
Dataset B \cite{zhuang2020smart} & $\checkmark$ & \ding{55} & \ding{55} & \ding{55} & 2 \\
\hline
Dataset C \cite{yu2023pscvfinder} & $\checkmark$ & \ding{55} & \ding{55} & \ding{55} & 3 \\
\hline
Dataset D \cite{qian2023cross} & $\checkmark$ & \ding{55} & \ding{55} & \ding{55} & 4 \\
\hline
iAudit \cite{ma2024combining} & $\checkmark$ & $\checkmark$ & $\checkmark$ & \ding{55} & 2* \\
\hline
Our Dataset & $\checkmark$ & $\checkmark$ & $\checkmark$ & $\checkmark$ & 5+** \\
\hline
\end{tabular}

\vspace{1mm}
\hspace{-3.9cm}\small 2* : iAudit focuses on logic and traditional vulnerabilities \\[2pt]
\hspace{-0.6cm}\small 5+** : includes 4 basic types and machine unauditable vulnerabilities (7 types)~\cite{zhang2023demystifying} \\[2pt]
\end{table}

\textbf{Motivation 1: Insufficient Quality and Comprehensiveness of Datasets.} We conducted a systematic evaluation of existing smart contract vulnerability datasets. As shown in Table \ref{dataset_comparison}, while Datasets A \cite{wu2021peculiar}, B \cite{zhuang2020smart}, C \cite{yu2023pscvfinder}, and D \cite{qian2023cross} provide vulnerability labels, they lack detailed vulnerability explanations and specific location information. Regarding vulnerability types, these datasets either have limited coverage, or like iAudit, focus mainly on logic vulnerabilities without detailed classification of traditional vulnerabilities. Notably, iAudit effectively leverages LLMs to provide vulnerability explanations and location information. While the positive samples are from verified audit reports, the negative samples and explanation enhancement rely on LLMs (including GPT-4 and GPT-3.5). Although constraints were added during explanation enhancement to ensure alignment with actual vulnerabilities, the process could benefit from more systematic verification by human experts. In contrast, our dataset not only utilizes multiple LLMs (Qwen2.5 and Mistral) for initial explanation generation but also introduces a rigorous expert review process. More importantly, by constructing expert-verified pairs of high-quality and suboptimal explanations (for example, for a reentrancy vulnerability, high-quality explanations detail the trigger conditions, attack paths, and specific impacts, while suboptimal explanations merely mention the presence of reentrancy risk or misunderstand critical state update sequences), our dataset can support large language models in DPO training, which is crucial for generating explanations aligned with human expert preferences. Our dataset not only comprehensively covers four major vulnerability types (RE, TD, IO, and DE) but also includes 7 types of MU vulnerabilities identified in recent research~\cite{zhang2023demystifying}.

\begin{figure}[htbp]
\centerline{\includegraphics[width=0.85\textwidth]{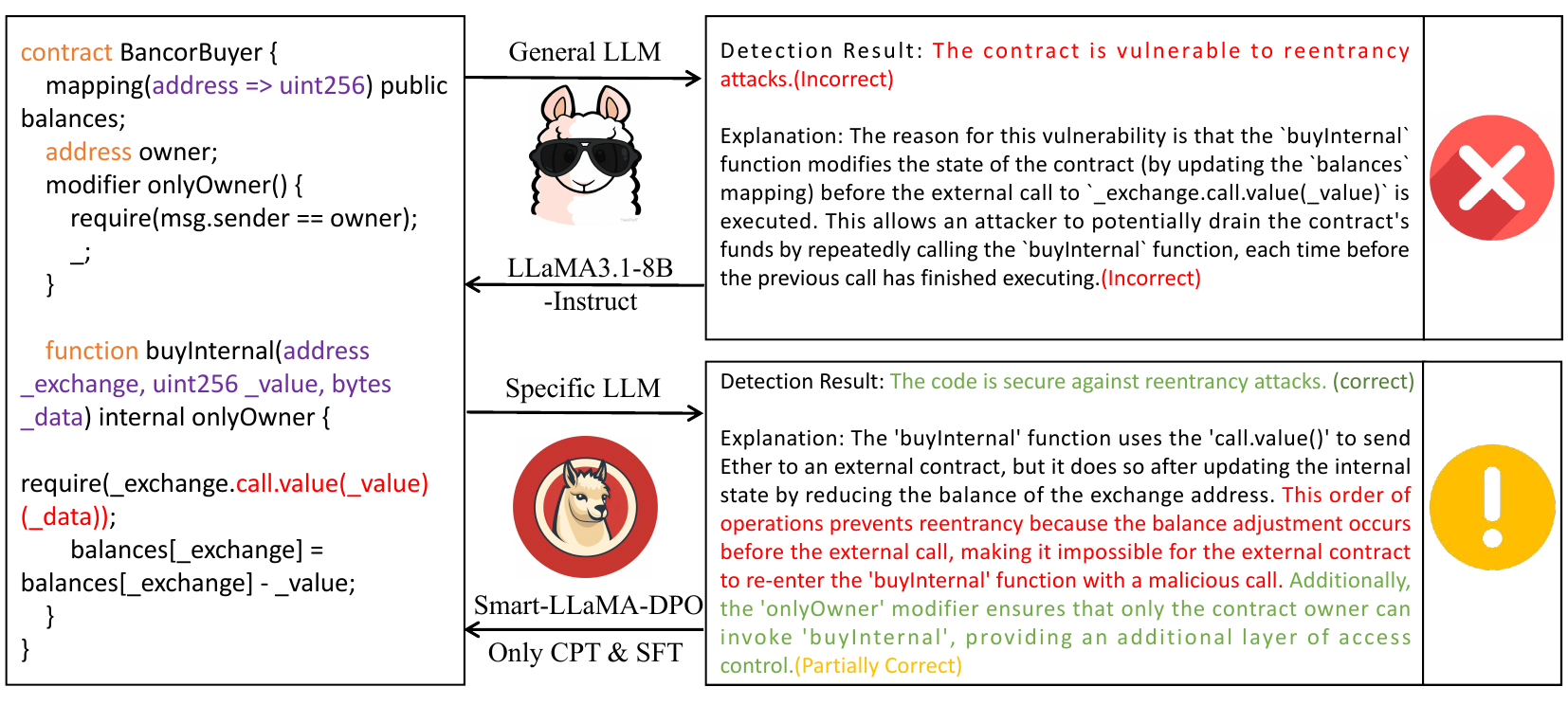}}
\caption{An motivation example to illustrate the limitations of LLM-based Explainable Smart Contract Vulnerability Detection, motivating the need for Direct Preference Optimization (DPO).}
\label{tab:motivation}
\end{figure}

\textbf{Motivation 2: Limitations in Detection Performance and Explanation Quality of LLM-based Smart Contract Vulnerability Detection Methods.} As shown in Fig. \ref{tab:motivation}, general LLMs (LLaMA3.1-8B-Instruct) fail in two aspects: they incorrectly identify a non-existent reentrancy vulnerability and provide inaccurate explanations. Although specialized LLMs like Smart-LLaMA-DPO (only trained through CPT and SFT) perform better in vulnerability detection, they still struggle to provide fully accurate explanations. The specialized LLM correctly identifies that the contract is safe against reentrancy attacks. However, its explanation contains a critical misunderstanding of the operation order. It erroneously states that the balance reduction occurs before the external call, whereas in the original code, the external call is executed first. This limitation in explanation quality persists despite the improvements from CPT and SFT. It underscores the need for further refinement in the LLM's understanding of smart contract vulnerability. Unlike traditional training methods that rely on absolute labels, DPO learns from relative preferences between paired explanations, which is crucial for smart contract security analysis. This approach is effective in two key aspects:

\begin{enumerate}
    \item \textbf{Fine-grained Security Pattern Learning:} DPO's loss function explicitly models the relative preference between two explanations, enabling the capture of subtle security-critical details in smart contracts. For instance, in reentrancy vulnerability detection, it can learn the critical distinction between explanations that correctly identify the ``checks-effects-interactions'' pattern versus those that misunderstand execution order. This is crucial as minor execution order misunderstandings can cause catastrophic security breaches in smart contracts.
    
    \item \textbf{Context-dependent Quality Assessment:} Smart contracts often contain complex interactions between multiple functions and state variables, making the quality of explanations highly context-dependent. DPO's preference learning mechanism is effective because it can capture how explanation quality varies with context via paired comparisons. For example, consider two explanations for a function containing both external calls and state updates:

    Explanation A: "This function updates the balance before making external calls."
    
    Explanation B: "This function makes external calls after completing all internal state changes, following the checks-effects-interactions pattern."

    While both explanations describe the same code behavior, Explanation B would be preferred in a reentrancy vulnerability analysis context as it explicitly links the ordering to a security pattern. However, for analyzing integer overflow vulnerabilities in the same function, Explanation A's focus on balance updates might be more relevant.

\end{enumerate}

Through this preference learning mechanism, DPO effectively addresses the limitations observed in models trained only with CPT and SFT. For instance, it helps prevent misunderstandings about operation ordering in reentrancy detection which is shown in Fig. \ref{tab:motivation}.

\section{Approach}

\begin{figure*}[htbp]
\centerline{\includegraphics[width=0.9\textwidth]{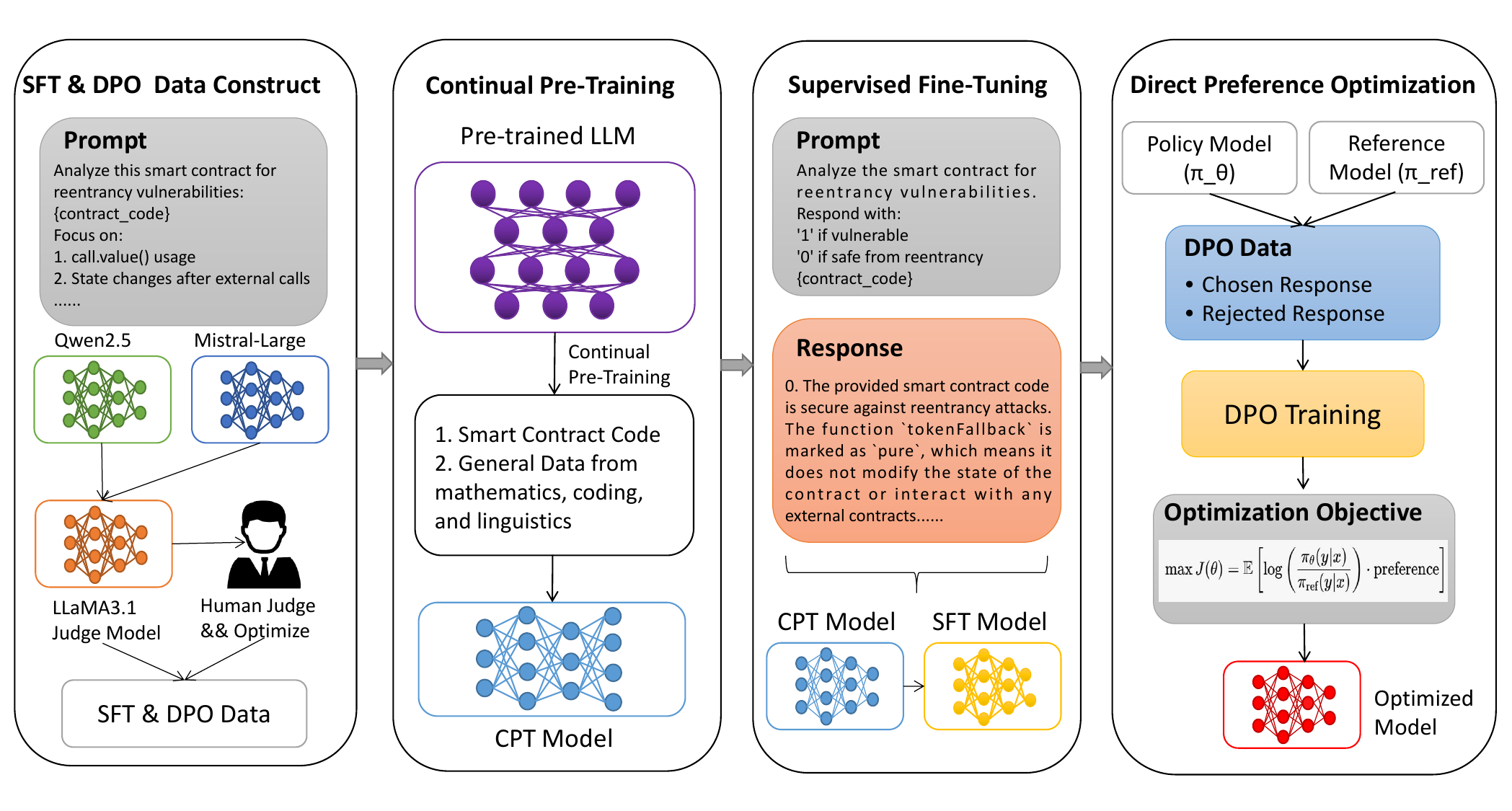}}
\caption{The Overview of our Smart-LLaMA-DPO.}
\label{overview}
\end{figure*}

Our Smart-LLaMA-DPO approach consists of four key stages illustrated in Fig. \ref{overview}: SFT and DPO Data Construct, Continual Pre-Training, Supervised Fine-Tuning, and Direct Preference Optimization. The process is supported by open-source smart contract vulnerability data collection and evaluation of explanations. We chose LLaMA-3.1-8B as the base model for its open-source nature, efficient parameter size and strong fine-tuning potential \cite{alrashedy2023language, lu2023llama}. We would like to point out that compared to iAudit \cite{ma2024combining}, Smart-LLaMA \cite{yu2024smart}, and FTSmartAudit \cite{wei2024leveraging} which focus primarily on optimizing detection performance while providing explanations, we additionally emphasize explanation quality through Direct Preference Optimization. Furthermore, compared to Smart-LLaMA \cite{yu2024smart}, our approach evaluates a broader range of vulnerability types and vulnerability instances.

\subsection{Open-source Smart Contract Vulnerability Data Collection}

\subsubsection{Continual Pre-training}

We based our Continual Pre-training dataset on research from \cite{storhaug2023efficient}, which underwent extensive filtering and quality checks. The process involved using Google BigQuery to identify all Ethereum smart contract addresses with at least one transaction. The researchers then accessed Etherscan for the source code. For uniqueness, we employed the Jaccard Index \cite{allamanis2019adverse} for token-based similarity detection. Smart contracts were decomposed into core business logic, library code, and imported files. This approach removed commonly repeated code while preserving unique implementation logic. Following \cite{wang2018ccaligner, storhaug2023efficient}, we grouped contracts by filename and filtered them using a 0.9 similarity threshold, eliminating contracts with over 90\% token similarity.

\subsubsection{Supervised Fine-Tuning and Direct Preference Optimization}

For Supervised Fine-Tuning and Direct Preference Optimization, we utilized labeled datasets from \cite{liu2023rethinking}, \cite{yu2023pscvfinder} and \cite{zhang2023demystifying}, covering various smart contract vulnerability types. The smart contract in \cite{liu2023rethinking} were manually verified for accuracy. \cite{yu2023pscvfinder} is based on the SmartBugs dataset \cite{ferreira2020smartbugs}, with reentrancy vulnerabilities annotated by \cite{wu2021peculiar}. Zhang et al. \cite{zhang2023demystifying} collects machine unauditable vulnerabilities from the highly reputable Code4rena contests. We selected 1,634 smart contracts containing \texttt{call.value} from \cite{yu2023pscvfinder} and identified 379 contracts with \texttt{delegatecall}, which we manually labeled. To expand our dataset, we collected verified contracts from Etherscan, GitHub repositories, and blog posts, all dated 2020-2024. 

\subsection{SFT and DPO Data Construct}

\subsubsection{Initial Data Generation}

We designed a multi-stage data generation process for our study, built upon recent work \cite{yu2024smart, wang2022self}. We utilized Qwen2.5-72B-Instruct and Mistral-Large-Instruct-2407-123B as our initial models to generate detailed explanations of smart contract vulnerabilities. These open-source LLMs were chosen for their powerful natural language processing and code comprehension capabilities, comparable to GPT-4 Turbo but at a lower cost \cite{yang2024qwen2, qwen2.5, mistrallarge2, sun2024llm4vuln}.

We designed specialized prompts for each vulnerability type (RE, TD, DE, IO, and MU), implementing label-guided analysis to help the LLM explain vulnerabilities and pinpoint their exact locations in the code.

For each vulnerability type, the prompts were tailored to focus on specific aspects:

\begin{itemize}
    \item RE: Use of call.value(), operation order, external calls, access control, and internal function implementation.
    \item TD: Use of block.timestamp or now, time constraints in critical operations, potential miner manipulation, and precision of time measurements impacting contract logic.
    \item DE: Use of delegatecall(), context preservation, state variable manipulation, access control, and internal function implementation.
    \item IO: Arithmetic operations (especially on uint variables), use of SafeMath library or Solidity 0.8.x built-in overflow/underflow checks, use of the 'unchecked' keyword (Solidity 0.8.x or higher), arithmetic operations in critical operations, and type conversion.
    \item MU: Based on audit report analysis, prompts guide the LLM to identify vulnerabilities identified in \cite{zhang2023demystifying}, explaining their root causes and exploit mechanisms in detail.
\end{itemize}

\subsubsection{LLM Scoring}
\label{LLM_socoring}

We employed Llama-3.1-70B-Instruct (validated in Section~\ref{subsubsec:llm_evaluation}) as an evaluation model to assess explanations from Qwen2.5 and Mistral based on three criteria: correctness (0.6 weight), thoroughness (0.3 weight), and clarity (0.1 weight), each on a 1-10 scale. Correctness evaluates vulnerability identification accuracy and reasoning logic. Thoroughness assesses coverage of vulnerability issues. Clarity measures explanation structure and applicability. Explanations with the highest weighted composite scores (WCS = 0.6×correctness + 0.3×thoroughness + 0.1×clarity) were selected for human review.

\subsubsection{SFT Data Construction}

To minimize potential bias, we selected 8 senior PhD students specializing in blockchain: 4 with extensive smart contract security auditing experience and 4 with academic publications in top conferences. They were divided into 4 balanced groups, each containing one industry expert and one academic expert to review one specific vulnerability type. To ensure consistent vulnerability labeling, we established specific annotation guidelines: For \textbf{RE}, our labeling criteria cover execution order analysis between external calls and state variable updates, reentrancy risk assessment standards in cross-contract invocation scenarios, identification of hidden reentrant call points in contract inheritance (both ETH and token transfers), and evaluation of reentrancy guard effectiveness. For \textbf{TD}, we examine block.timestamp's indirect impacts on contract state transitions. For randomness-related scenarios, we evaluate specific risks of timestamp-based random number generation in applications like gaming and lottery systems. For \textbf{IO}, we consider analysis of multiple overflow possibilities in complex expressions and overflow risks from type conversions (e.g., uint8 to uint256). The annotation covers implementation requirements for overflow checking mechanisms across different Solidity versions and potential issues from improper 'SafeMath' usage. For \textbf{DE}, we refine storage layout analysis standards in proxy patterns. For contract upgrade mechanisms, we include access control defect identification criteria, verification methods for callee contract code integrity, and security risk assessment standards across different upgrade scenarios. For \textbf{MU}, we verify whether the LLM-generated explanations align with the corresponding audit reports. Any discrepancies are corrected to strictly match the reports. Experts reviewed high-scoring LLM explanations and made necessary corrections. To ensure review consistency, when significant disagreements arise between the two reviewers within a group, a third expert from another group is selected to arbitrate. The final modifications are determined through a three-way discussion. To address potential bias, we introduced an external review stage after the initial annotation. Two senior blockchain and smart contract developers with extensive experience reviewed the annotated explanations which ensured broader perspectives.

\subsubsection{DPO Data Construction}

The DPO dataset consists of pairs of preferred and rejected outputs. Preferred outputs are taken from the high-quality outputs in the SFT dataset, following the annotation guidelines detailed in the SFT phase. For rejected outputs, experts rewrite a "suboptimal" version based on the lower-scoring outputs from LLaMA3.1. These suboptimal versions maintain basic correctness while intentionally reducing analysis depth: for \textbf{RE}, they only identify obvious external calls without analyzing execution ordering, inheritance-based reentrant risks, or token transfer scenarios; for \textbf{TD}, they merely mark direct block.timestamp usage without examining state transition impacts or analyzing timestamp-based randomness in gaming/lottery applications; for \textbf{IO}, they only flag simple overflow points without considering complex expressions, type conversions, or version-specific checking requirements; for \textbf{DE}, they simply identify basic DELEGATECALL usage without analyzing storage layout in proxy patterns or upgrade mechanism security. For \textbf{MU}, rejected outputs simplify or omit key details, such as audit findings or critical reasons. Experts ensure a noticeable but reasonable quality gap between preferred and rejected outputs. 

\subsection{Continual Pre-Training}

In the Continual Pre-Training phase, we focus on enhancing the LLM's comprehension of smart contract security. This process is guided by a domain-specific language modeling objective:

\begin{equation}
L_{CPT} = -\mathbb{E}{x \sim D}[\sum{i=1}^{n} \log P(x_i | x_{<i}, c_i)]
\end{equation}

Here, $D$ represents our curated dataset of smart contracts, $x$ is a sequence of tokens from a contract, $c_i$ denotes the contextual information surrounding $x_i$, and $P(x_i | x_{<i}, c_i)$ is the LLM's predicted probability of token $x_i$ given its preceding tokens and context.

This formulation encourages the LLM to learn key aspects: (1) Solidity syntax and semantics (keywords like 'function', 'uint256', 'require') for contract comprehension. (2) Design patterns like 'Checks-Effects-Interactions' to prevent reentrancy attacks, along with other security best practices, helping the LLM identify vulnerabilities. (3) Security-critical functions ('transfer', 'send', 'call.value()'), crucial for fund transfers and inter-contract interactions, allowing the LLM to recognize their correct usage and risks. (4) Contextual analysis skills to detect vulnerabilities by understanding contract structures, function interactions, and state variables. To maintain general capabilities and avoid catastrophic forgetting, diverse data from mathematics, coding, and linguistics is incorporated, enhancing its ability to generalize to novel contract patterns.

\subsection{Supervised Fine-Tuning}

The Supervised Fine-Tuning phase focuses on optimizing the LLM for vulnerability detection and explanation generation. We employ a balanced objective function:

\begin{equation}
L_{SFT} = \frac{1}{2}(L_{detect} + L_{explain})
\end{equation}

where:
\begin{equation}
L_{detect} = -\sum_{(x,y) \in D_{vul}} \log P(y|x;\theta)
\end{equation}
\begin{equation}
L_{explain} = -\sum_{(x,e) \in D_{exp}} \sum_{i=1}^{|e|} \log P(e_i|x, e_{<i};\theta)
\end{equation}

In this formulation, $x$ denotes the input smart contract code, while $y$ and $e$ represent the target outputs for the detection task (vulnerability labels) and generation task (explanations), respectively. The parameter $\theta$ encompasses all trainable components of the LLM, including attention mechanisms, feedforward networks, and embedding layers.

This balanced loss function ensures that the LLM is effectively trained on both vulnerability detection and explanation generation tasks. The domain-specific knowledge acquired during the Continual Pre-Training stage serves as a crucial foundation for this fine-tuning phase. It enables the LLM to comprehend the intricate context and subtle nuances of smart contracts while simultaneously honing its ability to detect vulnerabilities and generate informative explanations.

\subsection{Direct Preference Optimization}

In the final stage, we employ Direct Preference Optimization (DPO) \cite{rafailov2023direct} to align the LLM's outputs with the preferences of human experts in smart contract vulnerability analysis. DPO offers a simplified approach to preference learning without explicit reward modeling or reinforcement learning.

The core of DPO is based on the insight that the optimal policy $\pi^*$ for a reward function $r^*$ under a KL-constrained optimization objective can be expressed as:

\begin{equation}
\pi^*(y|x) = \frac{1}{Z(x)} \pi_\text{ref}(y|x) \exp\left(\frac{1}{\beta} r^*(x,y)\right)
\end{equation}

where $Z(x)$ is a normalization factor (also known as the partition function), $\pi_\text{ref}$ is a reference policy, and $\beta$ is a temperature parameter. In our smart contract vulnerability detection scenario:

\begin{itemize}
\item $x$: represents the input smart contract code snippet
\item $y$: represents the LLM-generated vulnerability analysis and explanation
\item $\pi_\text{ref}$: is our reference policy, typically the initially fine-tuned model such as the Smart-LLaMA-DPO model trained in the SFT stage
\item $\pi^*$: is the final optimized model we aim to obtain, capable of generating vulnerability analyses that align with expert preferences
\end{itemize}

By rearranging this equation, we can express the reward function in terms of the optimal policy:

\begin{equation}
r^*(x,y) = \beta \log\frac{\pi^*(y|x)}{\pi_\text{ref}(y|x)} + \beta \log Z(x)
\end{equation}

This allows us to reformulate the Bradley-Terry preference model in terms of policies rather than rewards:

\begin{equation}
p^*(y_1 \succ y_2 | x) = \sigma\left(\beta \log\frac{\pi^*(y_1|x)}{\pi_\text{ref}(y_1|x)} - \beta \log\frac{\pi^*(y_2|x)}{\pi_\text{ref}(y_2|x)}\right)
\end{equation}

where $\sigma$ is the logistic function, and $y_1$ and $y_2$ represent two different vulnerability analysis results.

Based on this, we can define the DPO loss function:

\begin{equation}
\mathcal{L}_\text{DPO}(\pi_\theta; \pi_\text{ref}) = -\mathbb{E}_{(x,y_w,y_l)\sim \mathcal{D}}\left[\log \sigma\left(\beta \log\frac{\pi_\theta(y_w|x)}{\pi_\text{ref}(y_w|x)} - \beta \log\frac{\pi_\theta(y_l|x)}{\pi_\text{ref}(y_l|x)}\right)\right]
\end{equation}

where:
\begin{itemize}
\item $(x, y_w, y_l)$ are triples from our curated dataset $\mathcal{D}$, representing input, preferred output, and non-preferred output respectively
\item $x$: smart contract code snippet that serves as the input for vulnerability analysis and forms the context for the LLM's assessment
\item $y_w$: expert-preferred vulnerability analysis result, which represents the high-quality output that aligns with expert judgment and security best practices
\item $y_l$: expert non-preferred vulnerability analysis result, typically containing errors, omissions, or less comprehensive explanations compared to $y_w$
\item $\pi_\theta$: our policy model being optimized, initialized as $\pi_\text{ref}$, which learns to generate outputs that align with expert preferences through the training process
\end{itemize}

This loss function encourages the LLM to assign higher probability to preferred outputs ($y_w$) compared to non-preferred outputs ($y_l$), effectively learning from expert preferences. By minimizing this loss, we directly optimize our policy model $\pi_\theta$ to align with the preferences of smart contract security experts, without the need for a separate reward model. 

\subsection{Evaluation of Explanations}

To comprehensively evaluate the quality of vulnerability explanations, we developed an evaluation framework inspired by \cite{wang2023generating, yu2024smart}, focusing on three aspects: correctness, thoroughness, and clarity. Each aspect is rated using a 4-point Likert scale \cite{joshi2015likert}. It differs from the 10-point scoring system used in dataset annotation (Section \ref{LLM_socoring}). The 10-point system offers fine granularity for curating high-quality explanations, while the 4-point Likert scale emphasizes simplicity and practicality.

\subsubsection{Evaluation Criteria}
\label{subsubsec:evaluation_criteria}

Our evaluation focuses on three core aspects using a 4-point scale (1-Strongly disagree to 4-Strongly agree):

\textbf{Correctness}: Measures logical reasoning and vulnerability identification accuracy. Scores range from 1 (significant errors in logic and identification) to 4 (correct logic with precise identification and localization).

\textbf{Thoroughness}: Assesses comprehensive coverage of all potential vulnerability points. Scores range from 1 (omits key vulnerabilities) to 4 (comprehensive coverage with detailed explanations).

\textbf{Clarity}: Evaluates whether explanations are clear, concise, and applicable. Scores range from 1 (verbose with unclear key points) to 4 (precise, lucid, directly applicable, no redundancy).

\subsubsection{LLM Evaluation}
\label{subsubsec:llm_evaluation}

We employed Llama-3.1-70B-Instruct for automated assessment, selected after comparing multiple LLMs on 40 explanation samples (8 per vulnerability type). Llama-3.1-70B-Instruct achieved 90\% agreement with human expert judgments, matching GPT-4 Turbo and outperforming GPT-4o (87.5\%), Qwen2.5-72B-Instruct (85\%), and GPT-3.5-Turbo (82.5\%), while being more cost-effective. Using carefully crafted prompts with scoring guidelines, the LLM evaluated explanations on correctness, thoroughness, and clarity using a 4-point Likert scale \cite{joshi2015likert}.

\subsubsection{Human Evaluation}

To further validate the evaluation results, we enlisted five experienced smart contract security experts for human evaluation. Each expert dedicated 8 hours to assessing explanations for one vulnerability category, totaling 40 hours. The experts utilized the same 4-point Likert scale \cite{joshi2015likert} as the LLM, scoring each explanation on correctness, thoroughness, and clarity. They also provided detailed scoring rationales and overall quality assessments. 

\section{Experiments}
\subsection{Research Questions}

To evaluate our Smart-LLaMA-DPO approach, we examine the following research questions:

$\bullet$ $\textbf{RQ1}$: How does Smart-LLaMA-DPO perform in detecting four key smart contract vulnerabilities compared to state-of-the-art methods?

$\bullet$ $\textbf{RQ2}$: Can Smart-LLaMA-DPO effectively detect machine-unauditable vulnerabilities that traditionally require manual analysis?

$\bullet$ $\textbf{RQ3}$: How do individual modules and Chain of Thought (COT) reasoning affect Smart-LLaMA-DPO's effectiveness?

$\bullet$ $\textbf{RQ4}$: How effective are the explanations generated by Smart-LLaMA-DPO in terms of correctness, thoroughness, and clarity?

$\bullet$ $\textbf{RQ5}$: How does Smart-LLaMA-DPO compare to existing methods in terms of practical applicability in real-world scenarios?

\subsection{Dataset}

\textbf{Continual Pre-training:} We employ a dataset derived from the work of Storhaug et al. \cite{storhaug2023efficient}. This dataset comprises 186,397 unique smart contract instances from the Ethereum blockchain, totalling 501.62M tokens. We also augment this dataset with an additional 100,000 instances from various domains, including general code, mathematics, English, and Chinese text, totalling 118.94M tokens. This results in a comprehensive dataset of 286,397 instances, totaling 620.56M tokens.

\textbf{Supervised Fine-Tuning:} Our Supervised Fine-Tuning dataset is curated from multiple sources, primarily drawing from Liu et al. \cite{liu2023rethinking} and Yu et al. \cite{yu2023pscvfinder}. We incorporate manually verified vulnerabilities from \cite{liu2023rethinking} and the SmartBugs dataset \cite{ferreira2020smartbugs}, with RE vulnerabilities annotated by Wu et al. \cite{wu2021peculiar}. Additionally, the dataset includes MU vulnerabilities extracted from Zhang et al. \cite{zhang2023demystifying}. To enrich the dataset, we manually annotated contracts from \cite{yu2023pscvfinder}, Etherscan, GitHub repositories, and blog posts, all dated 2020-2024. For Etherscan-sourced contracts specifically, we focused on token-level RE vulnerabilities. All datasets for SFT and DPO adhere to strict selection criteria: they are peer-reviewed and verified real-world contracts. The final dataset comprises 3,390 RE, 1,167 TD, 1,013 IO, 698 delegatecall, and 1281 MU instances, totalling 8.90M tokens.

\textbf{Direct Preference Optimization:} Our DPO dataset draws from the same sources as our SFT dataset, with no overlap between the two. The dataset contains 270 RE, 227 TD, 260 IO, 265 DE, and 420 MU instances, totaling 1.98M tokens.

\textbf{Evaluation:} Our evaluation dataset combines four vulnerability types from \cite{qian2023cross} (RE, TD, IO, DE) and machine-unauditable (MU) vulnerabilities from \cite{zhang2023demystifying}. Following \cite{qian2023cross}'s methodology, we sampled 20\% for RE, TD, and IO, while including all DE samples due to their limited number. We enhanced the dataset with ERC20 token RE cases and collected additional samples for all vulnerability types from EtherScan, GitHub repositories, and blog posts, all dated 2020-2024. We performed systematic cleaning to remove incorrect labels (e.g., cases mistakenly labeled as vulnerable where there were no state changes after call.value). The refined dataset contains 3,542 samples: RE (116/470, 13.27\%), TD (568/896, 25.30\%), IO (354/1,458, 41.16\%), DE (76/340, 9.60\%), and MU (116/378, 10.67\%). MU vulnerabilities include seven categories: PO (25/57), PE (15/46), IU (13/53), IS (23/59), EA (20/55), CI (10/55), and AV (10/53). Numbers in parentheses represent (vulnerable/total samples, percentage in evaluation dataset). The MU vulnerabilities are sourced from Zhang et al. \cite{zhang2023demystifying}, EtherScan, GitHub repositories, and blog posts, with no overlap between evaluation data and training data. For explanation quality evaluation, considering manual review costs, we maintained the same vulnerable/total samples ratios as in the refined dataset and randomly sampled approximately 200 samples per vulnerability type, resulting in 1,061 samples: RE (58/235), TD (142/224), IO (59/243), DE (38/170), and MU (58/189).

\subsection{Baselines}

Our evaluation includes a range of baselines for Smart Contract Vulnerability Detection, representing state-of-the-art approaches in four categories: rule-based, neural network-based, pre-trained model-based, and LLM-based techniques. \textbf{Rule-based techniques} include tools such as Mythril \cite{mueller2017mythril}, Osiris \cite{torres2018osiris}, Oyente \cite{luu2016making}, Slither \cite{feist2019slither}, Conkas\cite{veloso2021conkas}, Smartian \cite{choi2021smartian}, Confuzzius \cite{torres2021confuzzius}, sFuzz \cite{nguyen2020sfuzz}, Solhint \cite{protofire2020solhint}, Sailfish \cite{bose2022sailfish}, Securify \cite{tsankov2018securify}, and Smartcheck \cite{tikhomirov2018smartcheck}. \textbf{Neural network-based techniques} include GCN \cite{kipf2016semi}, TMP \cite{zhuang2020smart}, AME \cite{liu2021smart}, SMS \cite{qian2023cross} and DMT \cite{qian2023cross}. We faced challenges reproducing some neural network-based baselines, and in cases of significant discrepancies, we used the higher of the reproduced or reported results for fairness in comparisons. \textbf{Pre-trained models-based techniques}, rely on pre-trained models like CodeT5 \cite{wang2021codet5} and fine-tuning techniques to identify smart contract vulnerabilities, including Peculiar \cite{wu2021peculiar} and PSCVFinder \cite{yu2023pscvfinder}. \textbf{LLM-based techniques}, rely on LLMs to identify smart contract vulnerabilities, including Llama3.1-8B-Instruct \cite{dubey2024llama}, Llama3.1-70B-Instruct \cite{dubey2024llama}, Qwen2.5-7B-Instruct \cite{yang2024qwen2}, Qwen2.5-72B-Instruct \cite{yang2024qwen2}, GPT-4o (gpt-4o-2024-08-06) \cite{openai2024gpt4turbo}, Claude-3.5-Sonnet (claude-3-5-sonnet-20241022)  \cite{anthropic2024claude}, GPTScan \cite{sun2024gptscan}, GPTLens \cite{hu2023large}, and fine-tuning approaches such as FTSmartAudit \cite{wei2024leveraging} and iAudit \cite{ma2024combining}. All baselines were evaluated under consistent conditions using the same evaluation dataset. To ensure fairness, we fine-tuned FTSmartAudit \cite{wei2024leveraging} and iAudit \cite{ma2024combining} on their original datasets using their respective configurations but observed performance declines on our broader evaluation dataset. To address this, we fine-tuned on our training dataset to adapt them to the expanded evaluation scope.

\subsection{Metrics}

We evaluated our model performance from two dimensions: vulnerability identification capability and explanation quality. For vulnerability identification, we adopted four standard metrics: Precision measures the ratio of actual vulnerabilities among predicted positives, Recall indicates the proportion of detected vulnerabilities among all actual vulnerabilities, F1-score provides a balanced measure through the harmonic mean of Precision and Recall, and Accuracy reflects the overall correctness of predictions. For vulnerability explanation quality, we employed three key metrics: Correctness, Thoroughness, and Clarity, with detailed evaluation criteria specified in Section~\ref{subsubsec:evaluation_criteria}.

\begin{table}[!ht]
\setlength{\tabcolsep}{5pt} 
    \caption{Performance comparison with baselines (Part 1). Note: LLM-based techniques use Instruct versions of Llama3.1 and Qwen2.5. GPTScan and iAudit were originally designed primarily for logic vulnerabilities. Minor discrepancies (±0.01) between F1 and its calculation from precision/recall are due to rounding.}
    \label{TAB1a}
    \centering
    \begin{tabular}{@{}c@{\hspace{0pt}}|cccc@{\hspace{0pt}}|cccc@{}}
        \toprule
        \raisebox{-1\height}{\centering Methods} & \multicolumn{4}{c|}{Reentrancy} & \multicolumn{4}{c}{Timestamp Dependency}\\ 
        & A(\%) & P(\%) & R(\%) & F1(\%) & A(\%) & P(\%) & R(\%) & F1(\%)\\ 
        \midrule
        Mythril & 57.66 & 33.73 & 74.14 & 46.36 & 43.30 & 56.79 & 44.19& 49.70\\ 
        Osiris & 36.38 & 26.84 & 91.38 & 41.49 & 51.00 & 72.47 & 36.62 & 48.65\\ 
        Oyente & 70.85 & 42.22 & 49.14 & 45.42 & 42.52 & 66.67 & 18.66 & 29.16\\ 
        Slither & 42.13 & 16.09 & 31.90 & 21.39 &46.88 & 56.50 & 70.42 & 62.70\\ 
        Smartcheck & 48.30 & 30.09 & 82.76 & 44.14 & 39.51 & 57.30 & 17.96 & 27.35\\ 
        Conkas & 68.09 & 38.36 & 48.28 & 42.75 & 17.86 & 32.79 & 28.17 & 30.30\\
        Smartian & 53.62 & 27.03 & 51.72 & 35.50 & 23.66 & 39.26 & 37.32 & 38.27\\
        Confuzzius & 74.04 & 48.05 & 63.79 & 54.81 & -- & -- & -- & --\\
        sFuzz & 48.09 & 10.98 & 15.52 & 12.86 & 16.96 & 31.67 & 26.76 & 29.01\\
        Solhint & 65.53 & 36.78 & 55.17 & 44.14 & 35.27 & 48.60 & 36.62 & 41.77\\
        Sailfish & 76.17 & 51.16 & 75.86 & 61.11 & -- & -- & -- & --\\
        Securify & 55.32 & 28.04 & 51.72 & 36.36 & -- & -- & -- & --\\
        \midrule
        GCN & 73.21 & 74.47 & 73.18 & 73.82 & 75.91 & 74.93 & 77.55 & 76.22\\ 
        TMP & 76.45 & 76.04 & 75.30 & 75.67 & 78.84 & 78.68 & 76.09 & 77.36\\ 
        AME & 81.06 & 79.62 & 78.45 & 79.03 & 82.25 & 81.42 & 80.26 & 80.84\\
        SMS & 83.85 & 79.46 & 77.48 & 78.46 & 89.77 & 89.15 & 91.09  & 90.11\\ 
        DMT & 89.42 & 83.62 & 81.06 & 82.32 & 94.58 & 93.60 & 96.39 & 94.97\\ 
        \midrule
        Peculiar & 65.11 & 35.00 & 48.28 & 40.58 & 68.30 & 77.10 & 71.13 & 73.99\\ 
        PSCVFinder & 64.68 & 34.94 & 50.00 & 41.13 & 39.29 & 52.17 & 50.70 & 51.43\\ 
        \midrule
        LLaMA3.1-8B & 33.19 & 25.13 & 86.21 & 38.91 & 56.25 & 68.64 & 57.04 & 62.31\\ 
        Qwen2.5-7B & 25.53 & 24.89 & 100.00 & 39.86 & 69.20 & 68.34 & 95.77 & 79.77\\ 
        LLaMA3.1-70B & 24.68 & 24.68 & 100.00 &39.59 & 63.39 & 63.39 & 100.00 & 77.60\\ 
        Qwen2.5-72B & 25.96 & 25.00 & 100.00 & 40.00 & 63.39 & 63.39 & 100.00 & 77.60\\
        GPT-4o & 57.45 & 34.78 & 82.76 & 48.98 & 49.55 & 80.85 & 26.76 & 40.21\\
        Claude-3.5-Sonnet & 26.38 & 25.11 & 100.00 & 40.14 & 70.09 & 85.05 & 64.08 & 73.09\\
        GPTScan & 32.77 & 25.96 & 93.10 & 40.60 & 60.27 & 62.56 & 92.96 & 74.79\\
        GPTLens & 27.23 & 25.33 & 100.00 & 40.42 & 62.50 & 63.06 & 98.59 & 76.92\\
        FTSmartAudit & 32.34 & 26.07 & 94.83 & 40.89 & 71.88 & 83.19 & 69.72 & 75.86\\
        iAudit & 72.77 & 46.88 & 77.59 & 58.44 & 62.50 & 63.81 & 94.37 & 76.14\\
        \textbf{Smart-LLaMA-DPO} & \textbf{94.47} & \textbf{90.91} & \textbf{86.21} & \textbf{88.50} & \textbf{95.54} & \textbf{97.83} & \textbf{95.07} & \textbf{96.43}\\
        \bottomrule
    \end{tabular}
    \label{performance_1}
\end{table}

\begin{table}[!ht]
\setlength{\tabcolsep}{5pt} 
    \caption{Performance comparison with baselines (Part 2). Note: LLM-based techniques use Instruct versions of Llama3.1 and Qwen2.5 models. GPTScan and iAudit were originally designed primarily for logic vulnerabilities. Minor discrepancies (±0.01) between F1 and its calculation from precision/recall are due to rounding.}
    \label{TAB1b}
    \centering
    \begin{tabular}{@{}c@{\hspace{0pt}}|cccc@{\hspace{0pt}}|cccc@{}}
        \toprule
        \raisebox{-1\height}{\centering Methods} & \multicolumn{4}{c|}{Overflow/Underflow} & \multicolumn{4}{c}{Delegatecall}\\ 
        & A(\%) & P(\%) & R(\%) & F1(\%) & A(\%) & P(\%) & R(\%) & F1(\%)\\ 
        \midrule
        Mythril & 42.46 & 20.25 & 46.61 & 28.23 & 61.76 & 33.73 & 73.68 & 46.28\\ 
        Osiris & 67.97 & 41.27 & 75.42 & 53.35 & -- & -- & -- & --\\ 
        Oyente & 77.78 & 55.00 & 46.61 & 50.46 & 67.65 & 30.23 & 34.21 & 32.10\\ 
        Slither & 56.93 & 27.01 & 45.48 & 33.89 & 52.65 & 31.11 & 92.11 & 46.51\\ 
        Smartcheck & 56.04 & 24.51 & 38.98  & 30.10 & 55.29 & 25.32 & 51.32 & 33.91\\ 
        Conkas & 51.85 & 22.12 & 38.98 & 28.22 & 75.29 & 47.22 & 89.47 & 61.82\\
        Smartian & 76.95 & 51.90 & 69.49 & 59.42 & 79.41 & 52.83 & 73.68 & 61.54\\
        Confuzzius & 79.84 & 58.33 & 59.32 & 58.82 & 72.35 & 41.82 & 60.53 & 49.46 \\
        sFuzz & 79.01 & 58.33 & 47.46 & 52.34 & 67.06 & 36.76 & 65.79 & 47.17\\
        Solhint & -- & -- & -- & -- & 69.41 & 40.54 & 78.95 & 53.57 \\
        \midrule
        GCN & 67.53 & 69.52 & 70.93 & 70.22 & 65.76 & 69.01 & 69.74 & 69.37\\ 
        TMP & 70.85 & 70.26 & 69.47 & 69.86 & 69.11 & 68.18 & 70.37 & 69.26\\ 
        AME & 73.24 & 71.36 & 71.59 & 71.47 & 72.85 & 70.25 & 69.40 & 69.82\\
        SMS & 79.36 & 78.14 & 72.98 & 75.47 & 78.82 & 76.97 & 73.69 & 75.29\\ 
        DMT & 85.64 & 85.44 & 74.32 & 79.49 & 82.76 & 84.61 & 77.93 & 81.13\\ 
        \midrule
        Peculiar & 82.72 & 61.64 & 76.27 & 68.18 & 84.12 & 65.71 & 60.53 & 63.01\\ 
        PSCVFinder & 64.20 & 33.72 & 49.15 & 40.00 & 90.59 & 76.19 & 84.21 & 80.00\\ 
        \midrule
        LLaMA3.1-8B & 54.32 & 30.00 & 66.10 & 41.27 & 65.29 & 34.33 & 60.53 & 43.81\\ 
        Qwen2.5-7B & 74.90 & 47.50 & 32.20 & 38.38 & 67.06 & 40.22 & 97.37 & 56.92\\ 
        LLaMA3.1-70B & 79.42 & 55.42 & 77.97 & 64.79 & 64.12 & 38.38 & 100.00 & 55.47\\ 
        Qwen2.5-72B & 81.96 & 65.02 & 55.65 & 59.97 & 68.24 & 41.30 & 100.00 & 58.46\\
        GPT-4o & 72.02 & 44.44 & 61.02 & 51.43 & 57.65 & 34.55 & 100.00 & 51.35\\
        Claude-3.5-Sonnet & 77.37 & 51.75 & 100.00 & 68.21 & 54.71 & 33.04 & 100.00 & 49.67\\
        GPTScan & 45.68 & 27.33 & 74.58 & 40.00 & 47.06 & 29.37 & 97.37 & 45.12\\
        GPTLens & 57.61 & 31.67 & 64.41 & 42.46 & 70.59 & 43.18 & 100.00 & 60.32\\
        FTSmartAudit & 79.42 & 59.18 & 49.15 & 53.70 & 56.47 & 32.69 & 89.47 & 47.89\\
        iAudit & 69.55 & 40.96 & 57.63 & 47.89 & 40.59 & 27.34 & 100.00 & 42.94\\
        \textbf{Smart-LLaMA-DPO} & \textbf{94.65} & \textbf{94.23} & \textbf{83.05} & \textbf{88.29} & \textbf{94.12} & \textbf{100.00} & \textbf{73.68} & \textbf{84.85}\\
        \bottomrule
    \end{tabular}
    \label{performance_2}
\end{table}
\subsection{Implementation Details}

We perform Continual Pre-training, Supervised Fine-Tuning and Direct Preference Optimization using LlamaFactory~\cite{zheng2024llamafactory} and DeepSpeed~\cite{rasley2020deepspeed} with \texttt{fp16} enabled.
We calculate loss with cross-entropy and optimize parameters using AdamW~\cite{adamw} with $\beta$=$(0.9, 0.99)$ and $\epsilon$=$1$e-$8$. For all our models, we employ full parameter tuning. During Continual Pre-training, we set the batch size to $64$ per device, gradient accumulation steps to $16$, epochs to $2$, learning rate to $1$e-$5$ with cosine decay, warmup steps to $0$, cutoff length to $2048$, and save steps to $500$. During Supervised Fine-Tuning, we set the batch size to $8$ per device, gradient accumulation steps to $8$, epochs to $3$, learning rate to $1$e-$5$ with cosine decay, warmup steps to $0$, cutoff length to $2048$, and save steps to $50$. For DPO training, we set the cutoff length to $1024$, batch size to $8$ per device, gradient accumulation steps to $1$, learning rate to $1$e-$5$ with cosine decay, epochs to $10$, and warmup steps to $0$. All models were trained on a server equipped with 8 NVIDIA H800 GPUs, each with 80GB memory. For evaluating our Smart-LLaMA-DPO, we use greedy decoding with do\_sample set to \texttt{false}.

\subsection{Experimental Results}

1) RQ1: To answer this question, we compared Smart-LLaMA-DPO's performance against baseline methods across four different vulnerability types. The results are presented in Table \ref{performance_1} and Table \ref{performance_2}.

Smart-LLaMA-DPO excelled across all vulnerability types, consistently outperforming state-of-the-art (SOTA) methods. For Reentrancy vulnerabilities, it achieved an F1-score of 88.50\% and accuracy of 94.47\%, surpassing DMT by 7.51\% in F1-score and 5.65\% in accuracy. In Timestamp Dependency detection, it led with an F1-score of 96.43\% and accuracy of 95.54\%, outperforming DMT by 1.54\% and 1.02\%, respectively. For Overflow/Underflow vulnerabilities, it maintained the top position with an F1-score of 88.29\% and accuracy of 94.65\%, exceeding DMT by 11.07\% and 10.52\%. Finally, for Delegatecall vulnerabilities, it achieved an F1-score of 84.85\% and accuracy of 94.12\%, surpassing PSCVFinder by 6.06\% in F1-score and 3.90\% in accuracy. Compared to iAudit and FTSmartAudit, Smart-LLaMA-DPO benefits from the integration of DPO and CPT. iAudit's two-stage architecture separates detection and explanation tasks, which may affect its ability to leverage contextual information effectively. Meanwhile, FTSmartAudit, which lacks DPO, cannot achieve fine-grained preference learning. Notably, Smart-LLaMA-DPO consistently outperformed static analysis tools such as Mythril, Slither, Smartian, Solhint and Conkas, as well as LLM-based approaches including GPT-4o, Claude-3.5-Sonnet, GPTScan, and GPTLens across all metrics.

\vspace{5pt} 
\noindent\begin{tikzpicture}
  \node[draw=black, thick, fill=gray!20, rounded corners, inner sep=10pt, text width=0.92\linewidth] {
    \textbf{Answer to RQ1:} Smart-LLaMA-DPO consistently outperformed state-of-the-art baseline methods across all four types of vulnerabilities (reentrancy, timestamp dependency, integer overflow/underflow, and delegatecall).
  };
\end{tikzpicture}
\vspace{5pt} 

2) RQ2: We evaluate Smart-LLaMA-DPO's capability in detecting 7 types of machine-unauditable vulnerabilities identified in \cite{zhang2023demystifying}. Table~\ref{tab:unauditable} presents the comparative results with baseline methods. Smart-LLaMA-DPO demonstrates superior performance across all seven vulnerability types. For Price Oracle Manipulation, it achieves 87.7\% accuracy and 86.3\% F1-score, significantly surpassing iAudit. Notably, for Privilege Escalation, it achieves 91.3\% accuracy and 84.6\% F1-score, with improvements of 2.47\% and 3.80\% compared to iAudit. Even for complex vulnerabilities like Inconsistent State Updates, it maintains robust performance with 89.8\% accuracy and 85.0\% F1-score.

\begin{table}[htbp]
\caption{Performance Comparison on Machine-unauditable Vulnerabilities. LLaMA3.1-8B denotes LLaMA3.1-8B-Instruct. Total represents the overall performance across all MU vulnerabilities.}
\label{tab:unauditable}
\renewcommand{\arraystretch}{0.93}
\begin{tabular}{@{\hspace{0pt}}l|@{\hspace{3pt}}c@{\hspace{3pt}}|@{\hspace{3pt}}c@{\hspace{3pt}}|@{\hspace{3pt}}c@{\hspace{3pt}}|@{\hspace{3pt}}c@{\hspace{3pt}}|@{\hspace{3pt}}c@{\hspace{3pt}}|@{\hspace{3pt}}c@{\hspace{3pt}}|@{\hspace{3pt}}c@{\hspace{3pt}}|@{\hspace{3pt}}c@{\hspace{1pt}}}
\hline
\raisebox{-0.5\height}{\centering Method} & PO & EA & IU & IS & PE & AV & CI & Total \\
& Acc/F1 & Acc/F1 & Acc/F1 & Acc/F1 & Acc/F1 & Acc/F1 & Acc/F1 & Acc/F1 \\
\hline
LLaMA3.1-8B & 47.4/60.5 & 43.6/39.2 & 58.5/52.2 & 45.8/57.9 & 34.8/42.3 & 20.8/30.0 & 21.8/31.8 & 39.2/45.8 \\
GPT-4o & 49.1/62.3 & 54.6/59.0 & 58.5/50.0 & 69.5/71.0 & 67.4/66.7 & 49.1/27.0 & 58.2/41.0 & 57.9/56.4 \\
FTSmartAudit & 57.9/25.0 & 69.1/32.0 & 86.8/63.2 & 72.9/50.0 & 84.8/69.6 & 84.9/33.3 & 87.3/46.2 & 77.3/44.9 \\
iAudit & 40.4/55.3 & 52.7/55.2 & 88.7/72.7 & 88.1/82.9 & 89.1/81.5 & 88.7/62.5 & 92.7/80.0 & 76.7/66.2 \\
Ours & 87.7/86.3 & 87.3/80.0 & 92.5/83.3 & 89.8/85.0 & 91.3/84.6 & 92.5/77.8 & 94.6/82.4 & 90.7/83.4 \\
\hline
\end{tabular}
\end{table}

\noindent\begin{tikzpicture}
  \node[draw=black, thick, fill=gray!20, rounded corners, inner sep=10pt, text width=0.92\linewidth] {
    \textbf{Answer to RQ2:} Smart-LLaMA-DPO outperforms all baselines in detecting 7 types of machine-unauditable vulnerabilities, achieving the highest accuracy and F1-score across all vulnerability types.
  };
\end{tikzpicture}
\vspace{5pt} 

3) RQ3: Our ablation study elucidates the crucial roles of DPO and CPT. To provide a feedback loop and enhance explanation transparency, we further incorporate Chain-of-Thought (CoT) reasoning in our analysis. Following LLM4Vuln \cite{sun2024llm4vuln}, we design our CoT prompt to first instruct LLMs to summarize the functionality of the smart contract, then analyze vulnerabilities by examining relevant security patterns based on the specific vulnerability type, and finally determine whether it is vulnerable with explanations. The base model represents the full performance, "Base+CoT" shows the base model with COT, "w/o dpo" indicates the model without DPO, "w/o cpt" represents the LLM without CPT, and "w/o dpo \& cpt" shows the performance without both components.

\begin{table}[htbp]
\centering
\caption{Ablation Study of Smart-LLaMA-DPO. MU represents machine-unauditable vulnerabilities.}
\begin{tabular}{|c|c|c|c|c|c|c|}
\hline
Types & Metric & Base & Base+CoT & w/o dpo & w/o cpt & w/o dpo \& cpt \\
\hline
RE & Acc(\%) & 94.47 & 94.47 & 83.40  & 86.38  & 90.21 \\
 & F1(\%) & 88.50 & 88.50 & 73.47 & 77.78  & 79.65 \\
\hline
TD & Acc(\%) & 95.54 & 93.75 & 80.80 & 82.14 & 69.64 \\
 & F1(\%) & 96.43 & 95.30 & 85.32 & 86.39 & 69.64 \\
\hline
IO & Acc(\%) & 94.65 & 93.42 & 89.71  & 83.95 & 85.19 \\
 & F1(\%) & 88.29 & 86.21 & 81.75 & 53.01 & 56.10 \\
\hline
DE & Acc(\%) & 94.12 & 94.12 & 93.53  & 93.53 & 91.76\\
 & F1(\%) & 84.85 & 84.85 & 83.08 & 83.08 & 78.12 \\
\hline
MU & Acc(\%) & 90.74 & 91.53 & 78.84 & 86.24 & 72.22 \\
 & F1(\%) & 83.41 & 85.19 & 71.22 & 80.60 & 66.88 \\
\hline
\end{tabular}
\label{tab:vulnerability_detection}
\end{table}

\textbf{For RE}, removing DPO training significantly reduces performance (accuracy drops from 94.47\% to 83.40\%, F1 from 88.50\% to 73.47\%), while removing CPT has a smaller impact (accuracy 86.38\%, F1 77.78\%). This shows that reentrancy detection relies more on DPO for fine-grained preference learning. This aligns with the need to distinguish subtle differences in the "Checks-Effects-Interactions" pattern, primarily learned through paired examples in DPO training. \textbf{For IO}, removing CPT causes a significant drop in performance (F1 drops from 88.29\% to 53.01\%), while removing DPO has minimal impact. Detection relies on understanding Solidity's type system and arithmetic rules, which is mainly acquired through CPT. While DPO improves explanation, core detection relies on CPT’s domain knowledge. \textbf{For TD}, both CPT and DPO have significant impacts. The full model achieves an F1 score of 96.43\%, but removing either component significantly degrades performance, and removing both drops accuracy to 69.64\%. This reflects the dual complexity of timestamp dependency vulnerabilities, which require CPT to capture block.timestamp semantics and DPO to optimize risk differentiation in time-dependent scenarios. \textbf{For DE}, CPT and DPO have relatively balanced roles. Detection requires CPT to understand EVM storage mechanics and DPO to identify storage layout mismatch risks, with either component's removal leading to moderate performance degradation. \textbf{For MU}, results reveal significant reliance on DPO training (removing DPO: accuracy drops from 90.74\% to 78.84\%, F1 from 83.41\% to 71.22\%), highlighting the importance of fine-grained preference learning. Removing CPT has a smaller impact (accuracy 86.24\%, F1 80.60\%), but still shows that domain-specific knowledge from pre-training enhances detection. The integration of CoT reasoning shows mixed effectiveness. For simpler vulnerabilities like TD and IO, performance drops (TD: -1.87\% accuracy, -1.17\% F1; IO: -1.30\% accuracy, -2.36\% F1), likely due to CoT overcomplicating these straightforward pattern recognition tasks. For DE and RE vulnerabilities, performance remains stable. However, for complex vulnerabilities like MU, CoT improves performance (+0.87\% accuracy, +2.13\% F1), highlighting its strength in analyzing complex logic and dependencies.

\vspace{5pt} 
\noindent\begin{tikzpicture}
  \node[draw=black, thick, fill=gray!20, rounded corners, inner sep=10pt, text width=0.92\linewidth] {
    \textbf{Answer to RQ3:} Our findings confirm that the combination of continual pre-training and DPO training creates strong complementary effects, effectively addressing the diverse requirements of different vulnerability types and achieving SOTA detection performance across all vulnerability types.
  };
\end{tikzpicture}
\vspace{5pt} 

4) RQ4: Table \ref{tab:ratings} presents the quality assessment results of explanations generated by the Smart-LLaMA-DPO and other baselines. The evaluation comprises two parts: LLM Evaluation and Human Evaluation, each assessing three dimensions: Correctness, Thoroughness, and Clarity. Our model significantly outperformed the best baseline (iAudit) in both LLM evaluation and human evaluation. In the LLM evaluation, when combining scores 4 (agree) and 3 (somewhat agree), our model achieved higher total positive ratings across all metrics: 86.62\% compared to the baseline's 75.12\% for correctness (834+85 and 713+84 respectively), 90.10\% compared to 75.12\% for thoroughness (731+225 and 586+211 respectively), and 97.46\% compared to 93.03\% for clarity (565+469 and 747+240 respectively). Similarly in human evaluation, our model maintained consistently higher combined positive ratings (scores 4 and 3): 81.15\% compared to the baseline's 70.22\% for correctness (646+215 and 388+357 respectively), 83.88\% compared to 72.76\% for thoroughness (544+346 and 188+584 respectively), and 94.63\% compared to 83.98\% for clarity (569+435 and 473+418 respectively). 

\vspace{5pt} 
\noindent\begin{tikzpicture}
  \node[draw=black, thick, fill=gray!20, rounded corners, inner sep=10pt, text width=0.92\linewidth] {
    \textbf{Answer to RQ4:} Smart-LLaMA-DPO has demonstrated its capability to generate explanations for smart contract vulnerability detection that are correct, thorough, and clear, as validated by both LLM Evaluation and Human Evaluation.
  };
\end{tikzpicture}
\vspace{5pt} 

\begin{table}[htbp]
\centering
\caption{Ratings of Correctness, Thoroughness, and Clarity (LLaMA3.1-8B refers to the Instruct version).}
\label{tab:ratings}
\begin{tabular}{|l|cccc|cccc|cccc|}
\hline
 & \multicolumn{4}{c|}{Correctness} & \multicolumn{4}{c|}{Thoroughness} & \multicolumn{4}{c|}{Clarity} \\
\cline{2-13}
& 1 & 2 & 3 & 4 & 1 & 2 & 3 & 4 & 1 & 2 & 3 & 4 \\
\hline
\multicolumn{13}{|c|}{LLM Evaluation} \\
\hline
LLaMA3.1-8B & 116 & 201 & 165 & 579 & 42  & 229 & 332 & 458 & 30 & 204 & 578 & 249 \\
FTSmartAudit & 101 & 234 & 161 & 565 & 53 & 167 & 351 & 490 & 41 & 165 & 454 & 401\\
iAudit & 135 & 129 & 84 & 713 & 48 & 216 & 211 & 586 & 27 & 47 & 240 & 747\\
Ours & 56 & 86 & 85 & 834 & 9 & 96 & 225 & 731 & 2 & 25 & 469 & 565 \\
\hline
\multicolumn{13}{|c|}{Human Evaluation} \\
\hline
LLaMA3.1-8B & 76 & 303 & 326 & 356 & 70 & 266 & 457 & 268 & 29 & 212 & 623 & 197 \\
FTSmartAudit & 127 & 234 & 352 & 348 & 126 & 184 & 523 & 228 & 28 & 165 & 482 & 386\\
iAudit & 61 & 255 & 357 & 388 & 48 & 241 & 584 & 188 & 27 & 143 & 418 & 473\\
Ours & 19 & 181 & 215 & 646 & 18 & 153 & 346 & 544 & 9 & 48 & 435 & 569 \\
\hline
\end{tabular}
\end{table}

\begin{figure}[!h]
\centerline{\includegraphics[width=1\textwidth]{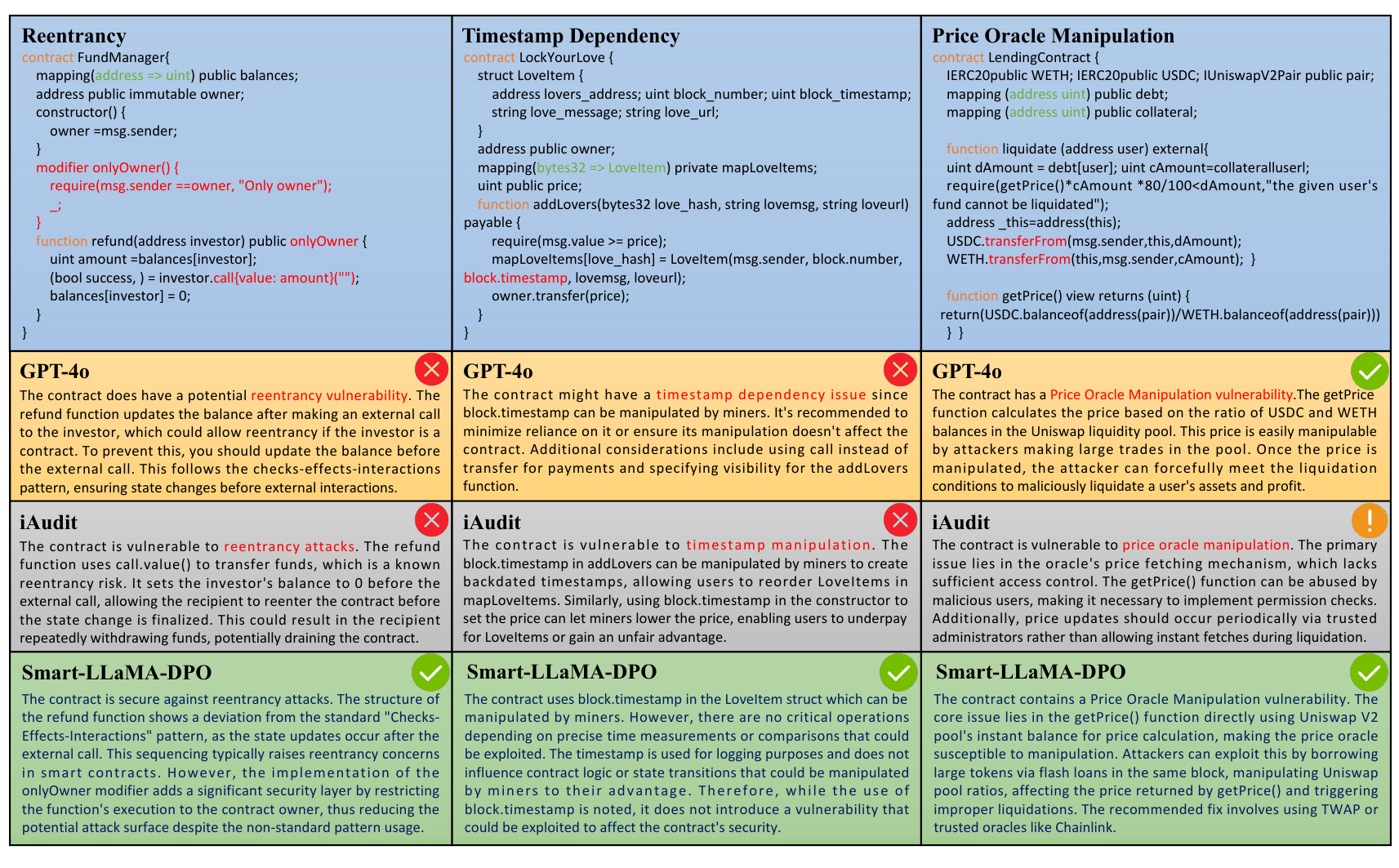}}
\caption{Case Study of Smart Contract Vulnerability Detection Using Smart-LLaMA-DPO.}
\label{smart_case}
\end{figure}

5) RQ5: We demonstrate the effectiveness of Smart-LLaMA-DPO in smart contract vulnerability detection through a comparative case study as shown in Fig. \ref{smart_case}. 

\textbf{Why we choose GPT-4o in GPT family and iAudit for this RQ.} GPT-4o (gpt-4o-2024-08-06) was chosen for its superior detection performance, with the highest average accuracy (58.91\%) and F1-score (49.67\%) across five vulnerability types, outperforming GPT-4 (accuracy: 56.49\%, F1-score: 48.50\%) and GPT-4 Turbo (accuracy: 50.61\%, F1-score: 49.54\%). Among works focusing on generating smart contract vulnerability explanations iAudit was selected for its higher average performance (accuracy: 64.42\%, F1-score: 58.32\%) compared to FTSmartAudit (accuracy: 63.48\%, F1-score: 52.65\%).

In the FundManager contract, GPT-4o identifies a reentrancy vulnerability risk, noting that the refund function updates the balance after an external call. iAudit also identifies a reentrancy vulnerability but incorrectly assumes that the balance is updated before the external call. In contrast, Smart-LLaMA-DPO provides more nuanced analysis, acknowledging the non-standard pattern usage while explaining how the onlyOwner modifier enhances security. For the LockYourLove contract, while GPT-4o and iAudit warn about potential block.timestamp manipulation by miners, Smart-LLaMA-DPO explains that the timestamp is only used for logging purposes and doesn't affect critical operations or contract logic, thus not constituting an actual vulnerability. For the LendingContract, GPT-4o and Smart-LLaMA-DPO correctly identify the root cause of the Price Oracle Manipulation vulnerability: reliance on the Uniswap pool's instant price. iAudit, however, focuses on secondary issues like access control permissions, reflecting how its two-stage architecture can lead to inconsistencies between detection and explanation.

\vspace{5pt} 
\noindent\begin{tikzpicture}
  \node[draw=black, thick, fill=gray!20, rounded corners, inner sep=10pt, text width=0.92\linewidth] {
    \textbf{Answer to RQ5:} Compared to GPT-4o and iAudit which produce false positives, Smart-LLaMA-DPO provides contextual security analysis by accurately assessing both potential risks and implemented security measures, enabling reliable distinction between theoretical vulnerabilities and actual security threats.
  };
\end{tikzpicture}

\section{Related Work}
\subsection{Smart Contract Vulnerability Detection}

Research in smart contract vulnerability detection has evolved significantly. Early approaches relied on program analysis techniques: some used symbolic execution to explore execution paths and identify vulnerabilities \cite{luu2016making}, while others employed pattern matching against known vulnerability signatures \cite{mueller2017mythril, tikhomirov2018smartcheck} or applied formal verification with custom compliance patterns \cite{tsankov2018securify}. The field then progressed to machine learning, particularly deep learning models, offering improved accuracy and generalization. Notable examples include bidirectional LSTMs with attention mechanisms for specific vulnerability detection \cite{qian2020towards}, and deep learning for measuring similarity to known vulnerable contracts \cite{gao2019smartembed}. Graph neural networks (GNNs) gained popularity due to their ability to capture contract structural information \cite{zhuang2020smart, liu2021smart}. Recent advancements include pre-trained models like Peculiar \cite{wu2021peculiar} for improved generalization, and innovative approaches such as prompt-tuning to bridge pre-training and downstream tasks \cite{yu2023pscvfinder}. Cross-modality learning approaches have also emerged, leveraging information from both bytecode and source code \cite{qian2023cross}. The latest frontier involves Large Language Models (LLMs). Studies by Chen et al. \cite{chen2023chatgpt} and David et al. \cite{david2023you} evaluated LLMs on real-world datasets, revealing both potential and challenges. Hu et al. \cite{hu2023large} explored LLMs' reasoning capabilities, while Sun et al. \cite{sun2024gptscan} introduced GPTScan for program analysis. FTSmartAudit \cite{wei2024leveraging} fine-tunes smaller LLMs for smart contract auditing, achieving comparable performance to state-of-the-art models. Ma et al. \cite{ma2024combining} proposed iAudit, a two-stage LLM framework for vulnerability detection and explanation, though its separated design may lead to inconsistencies.

\subsection{Reinforcement Learning in Software Engineering}

Reinforcement learning (RL) has gained increasing attention in software engineering in recent years \cite{abo2023role,nikanjam2022faults,chakraborty2024rlocator,reddy2020quickly,romdhana2022deep}. Several studies have explored RL applications in software tasks. Kim et al. \cite{kim2023adaptive} introduced ARAT-RL, an adaptive REST API testing technique using RL to prioritize operations. Corradini et al. \cite{corradini2024deeprest} proposed DeepREST for automated REST API testing using deep RL. In repository-level code completion, Wang et al. \cite{wang2024rlcoder} presented RLCoder for improving code generation. Li et al. \cite{li2024ircoco} developed IRCoCo, applying deep RL to code completion. More recently, Nashaat et al. \cite{nashaat2024towards} proposed the CodeMentor framework, which utilizes reinforcement learning with human feedback (RLHF) to optimize the performance of LLMs in software review tasks. However, the application of these techniques to explainable smart contract vulnerability detection remains unexplored. Our work applies preference-based optimization to this critical task. While not strictly reinforcement learning, our approach shares the goal of improving performance based on feedback.

\section{Threats to Validity}
\textbf{Internal Validity}: Our Smart-LLaMA-DPO occasionally produces redundant information in its outputs. We currently address this by truncating the generated text to retain only the initial portion. This issue likely stems from the reinforcement of certain patterns during the fine-tuning or DPO training process. To improve output quality, we are exploring advanced post-processing methods and refined post-training techniques for future iterations.

\textbf{External Validity}: The development of Smart-LLaMA-DPO relied heavily on a large corpus of annotated training data. Our data preparation involved a multi-step process using multiple LLMs and human verification. Despite these efforts, we cannot guarantee the absolute accuracy of every reasoning step in our training datasets. Furthermore, our current Smart-LLaMA-DPO's scope is limited to specific vulnerability types due to resource limitations. Future research will focus on developing more efficient data annotation methods and expanding our Smart-LLaMA-DPO's capability to address a wider array of smart contract vulnerabilities.
\section{Conclusion}

In this paper, we propose Smart-LLaMA-DPO, an advanced smart contract vulnerability detection approach based on LLMs. Unlike prior work, Smart-LLaMA-DPO utilizes a three-stage post-training process: continual pre-training, supervised fine-tuning, and direct preference optimization. In addition, we construct a comprehensive dataset covering four major vulnerability types and machine-unauditable vulnerabilities for both SFT and DPO. Our approach significantly outperforms state-of-the-art methods and can provide reliable explanations. While our current implementation only focuses on Solidity, Smart-LLaMA-DPO can be easily extended to other Domain-Specific Languages like Bash, SQL, and SysML by adapting the training corpus and annotation guidelines. 

\section{Data Availability}

All the source code and data are available at \cite{nansijie2025smartllamadpo}, and the trained model weights are available at \cite{smartllamadpo2025models}.

\section*{ACKNOWLEDGMENTS}
This work was supported by the National Key Research and Development Program of China (No.2023YFB3307203) and the Alliance of International Science Organizations Collaborative Research Program (No.ANSO-CR-KP-2022-03). 

\bibliographystyle{ACM-Reference-Format}
\bibliography{References}


\begin{thebibliography}{80}


\ifx \showCODEN    \undefined \def \showCODEN     #1{\unskip}     \fi
\ifx \showISBNx    \undefined \def \showISBNx     #1{\unskip}     \fi
\ifx \showISBNxiii \undefined \def \showISBNxiii  #1{\unskip}     \fi
\ifx \showISSN     \undefined \def \showISSN      #1{\unskip}     \fi
\ifx \showLCCN     \undefined \def \showLCCN      #1{\unskip}     \fi
\ifx \shownote     \undefined \def \shownote      #1{#1}          \fi
\ifx \showarticletitle \undefined \def \showarticletitle #1{#1}   \fi
\ifx \showURL      \undefined \def \showURL       {\relax}        \fi
\providecommand\bibfield[2]{#2}
\providecommand\bibinfo[2]{#2}
\providecommand\natexlab[1]{#1}
\providecommand\showeprint[2][]{arXiv:#2}

\bibitem[sma(2025)]%
        {smartllamadpo2025models}
 \bibinfo{year}{2025}\natexlab{}.
\newblock \bibinfo{title}{models of Smart-LLaMA-DPO}.
\newblock
\urldef\tempurl%
\url{https://doi.org/10.5281/zenodo.15255329}
\showURL{%
\tempurl}


\bibitem[nan(2025)]%
        {nansijie2025smartllamadpo}
 \bibinfo{year}{2025}\natexlab{}.
\newblock \bibinfo{title}{Smart-LLaMA-DPO}.
\newblock
\urldef\tempurl%
\url{https://gitlab.com/programmer-of-nansijie/smart-llama-dpo}
\showURL{%
\tempurl}


\bibitem[Abo-eleneen et~al\mbox{.}(2023)]%
        {abo2023role}
\bibfield{author}{\bibinfo{person}{Amr Abo-eleneen}, \bibinfo{person}{Ahammed Palliyali}, {and} \bibinfo{person}{Cagatay Catal}.} \bibinfo{year}{2023}\natexlab{}.
\newblock \showarticletitle{The role of Reinforcement Learning in software testing}.
\newblock \bibinfo{journal}{\emph{Information and Software Technology}} (\bibinfo{year}{2023}), \bibinfo{pages}{107325}.
\newblock


\bibitem[Alharby and Van~Moorsel(2017)]%
        {alharby2017blockchain}
\bibfield{author}{\bibinfo{person}{Maher Alharby} {and} \bibinfo{person}{Aad Van~Moorsel}.} \bibinfo{year}{2017}\natexlab{}.
\newblock \showarticletitle{Blockchain-based smart contracts: A systematic mapping study}.
\newblock \bibinfo{journal}{\emph{arXiv preprint arXiv:1710.06372}} (\bibinfo{year}{2017}).
\newblock


\bibitem[Allamanis(2019)]%
        {allamanis2019adverse}
\bibfield{author}{\bibinfo{person}{Miltiadis Allamanis}.} \bibinfo{year}{2019}\natexlab{}.
\newblock \showarticletitle{The adverse effects of code duplication in machine learning models of code}. In \bibinfo{booktitle}{\emph{Proceedings of the 2019 ACM SIGPLAN International Symposium on New Ideas, New Paradigms, and Reflections on Programming and Software}}. \bibinfo{pages}{143--153}.
\newblock


\bibitem[Alrashedy and Binjahlan(2023)]%
        {alrashedy2023language}
\bibfield{author}{\bibinfo{person}{Kamel Alrashedy} {and} \bibinfo{person}{Ahmed Binjahlan}.} \bibinfo{year}{2023}\natexlab{}.
\newblock \showarticletitle{Language Models are Better Bug Detector Through Code-Pair Classification}.
\newblock \bibinfo{journal}{\emph{arXiv preprint arXiv:2311.07957}} (\bibinfo{year}{2023}).
\newblock


\bibitem[Anthropic(2024)]%
        {anthropic2024claude}
\bibfield{author}{\bibinfo{person}{Anthropic}.} \bibinfo{year}{2024}\natexlab{}.
\newblock \bibinfo{title}{Claude-3.5-Sonnet}.
\newblock \bibinfo{howpublished}{\url{https://www.anthropic.com/claude}}.
\newblock


\bibitem[Authors(2024)]%
        {solidity_docs}
\bibfield{author}{\bibinfo{person}{The~Solidity Authors}.} \bibinfo{year}{2024}\natexlab{}.
\newblock \bibinfo{title}{Solidity Documentation}.
\newblock
\urldef\tempurl%
\url{https://docs.soliditylang.org/en/v0.8.28/}
\showURL{%
\tempurl}
\newblock
\shownote{Online documentation}.


\bibitem[Bose et~al\mbox{.}(2022)]%
        {bose2022sailfish}
\bibfield{author}{\bibinfo{person}{Priyanka Bose}, \bibinfo{person}{Dipanjan Das}, \bibinfo{person}{Yanju Chen}, \bibinfo{person}{Yu Feng}, \bibinfo{person}{Christopher Kruegel}, {and} \bibinfo{person}{Giovanni Vigna}.} \bibinfo{year}{2022}\natexlab{}.
\newblock \showarticletitle{Sailfish: Vetting smart contract state-inconsistency bugs in seconds}. In \bibinfo{booktitle}{\emph{2022 IEEE Symposium on Security and Privacy (SP)}}. IEEE, \bibinfo{pages}{161--178}.
\newblock


\bibitem[Chakraborty et~al\mbox{.}(2024)]%
        {chakraborty2024rlocator}
\bibfield{author}{\bibinfo{person}{Partha Chakraborty}, \bibinfo{person}{Mahmoud Alfadel}, {and} \bibinfo{person}{Meiyappan Nagappan}.} \bibinfo{year}{2024}\natexlab{}.
\newblock \showarticletitle{RLocator: Reinforcement learning for bug localization}.
\newblock \bibinfo{journal}{\emph{IEEE Transactions on Software Engineering}} (\bibinfo{year}{2024}).
\newblock


\bibitem[Chen et~al\mbox{.}(2023)]%
        {chen2023chatgpt}
\bibfield{author}{\bibinfo{person}{Chong Chen}, \bibinfo{person}{Jianzhong Su}, \bibinfo{person}{Jiachi Chen}, \bibinfo{person}{Yanlin Wang}, \bibinfo{person}{Tingting Bi}, \bibinfo{person}{Yanli Wang}, \bibinfo{person}{Xingwei Lin}, \bibinfo{person}{Ting Chen}, {and} \bibinfo{person}{Zibin Zheng}.} \bibinfo{year}{2023}\natexlab{}.
\newblock \showarticletitle{When chatgpt meets smart contract vulnerability detection: How far are we?}
\newblock \bibinfo{journal}{\emph{arXiv preprint arXiv:2309.05520}} (\bibinfo{year}{2023}).
\newblock


\bibitem[Chen et~al\mbox{.}(2020)]%
        {chen2020survey}
\bibfield{author}{\bibinfo{person}{Huashan Chen}, \bibinfo{person}{Marcus Pendleton}, \bibinfo{person}{Laurent Njilla}, {and} \bibinfo{person}{Shouhuai Xu}.} \bibinfo{year}{2020}\natexlab{}.
\newblock \showarticletitle{A survey on ethereum systems security: Vulnerabilities, attacks, and defenses}.
\newblock \bibinfo{journal}{\emph{ACM Computing Surveys (CSUR)}} \bibinfo{volume}{53}, \bibinfo{number}{3} (\bibinfo{year}{2020}), \bibinfo{pages}{1--43}.
\newblock


\bibitem[Choi et~al\mbox{.}(2021)]%
        {choi2021smartian}
\bibfield{author}{\bibinfo{person}{Jaeseung Choi}, \bibinfo{person}{Doyeon Kim}, \bibinfo{person}{Soomin Kim}, \bibinfo{person}{Gustavo Grieco}, \bibinfo{person}{Alex Groce}, {and} \bibinfo{person}{Sang~Kil Cha}.} \bibinfo{year}{2021}\natexlab{}.
\newblock \showarticletitle{Smartian: Enhancing smart contract fuzzing with static and dynamic data-flow analyses}. In \bibinfo{booktitle}{\emph{2021 36th IEEE/ACM International Conference on Automated Software Engineering (ASE)}}. IEEE, \bibinfo{pages}{227--239}.
\newblock


\bibitem[Corradini et~al\mbox{.}(2024)]%
        {corradini2024deeprest}
\bibfield{author}{\bibinfo{person}{Davide Corradini}, \bibinfo{person}{Zeno Montolli}, \bibinfo{person}{Michele Pasqua}, {and} \bibinfo{person}{Mariano Ceccato}.} \bibinfo{year}{2024}\natexlab{}.
\newblock \showarticletitle{DeepREST: Automated Test Case Generation for REST APIs Exploiting Deep Reinforcement Learning}.
\newblock \bibinfo{journal}{\emph{arXiv preprint arXiv:2408.08594}} (\bibinfo{year}{2024}).
\newblock


\bibitem[David et~al\mbox{.}(2023)]%
        {david2023you}
\bibfield{author}{\bibinfo{person}{Isaac David}, \bibinfo{person}{Liyi Zhou}, \bibinfo{person}{Kaihua Qin}, \bibinfo{person}{Dawn Song}, \bibinfo{person}{Lorenzo Cavallaro}, {and} \bibinfo{person}{Arthur Gervais}.} \bibinfo{year}{2023}\natexlab{}.
\newblock \showarticletitle{Do you still need a manual smart contract audit?}
\newblock \bibinfo{journal}{\emph{arXiv preprint arXiv:2306.12338}} (\bibinfo{year}{2023}).
\newblock


\bibitem[Dhillon et~al\mbox{.}(2017)]%
        {dhillon2017dao}
\bibfield{author}{\bibinfo{person}{Vikram Dhillon}, \bibinfo{person}{David Metcalf}, \bibinfo{person}{Max Hooper}, \bibinfo{person}{Vikram Dhillon}, \bibinfo{person}{David Metcalf}, {and} \bibinfo{person}{Max Hooper}.} \bibinfo{year}{2017}\natexlab{}.
\newblock \showarticletitle{The DAO hacked}.
\newblock \bibinfo{journal}{\emph{blockchain enabled applications: Understand the blockchain Ecosystem and How to Make it work for you}} (\bibinfo{year}{2017}), \bibinfo{pages}{67--78}.
\newblock


\bibitem[Dubey et~al\mbox{.}(2024)]%
        {dubey2024llama}
\bibfield{author}{\bibinfo{person}{Abhimanyu Dubey}, \bibinfo{person}{Abhinav Jauhri}, \bibinfo{person}{Abhinav Pandey}, \bibinfo{person}{Abhishek Kadian}, \bibinfo{person}{Ahmad Al-Dahle}, \bibinfo{person}{Aiesha Letman}, \bibinfo{person}{Akhil Mathur}, \bibinfo{person}{Alan Schelten}, \bibinfo{person}{Amy Yang}, \bibinfo{person}{Angela Fan}, {et~al\mbox{.}}} \bibinfo{year}{2024}\natexlab{}.
\newblock \showarticletitle{The llama 3 herd of models}.
\newblock \bibinfo{journal}{\emph{arXiv preprint arXiv:2407.21783}} (\bibinfo{year}{2024}).
\newblock


\bibitem[{Ethereum Foundation}(2024a)]%
        {solidity_timestamp}
\bibfield{author}{\bibinfo{person}{{Ethereum Foundation}}.} \bibinfo{year}{2024}\natexlab{a}.
\newblock \bibinfo{title}{Block and Transaction Properties}.
\newblock \bibinfo{howpublished}{\url{https://docs.soliditylang.org/en/latest/units-and-global-variables.html\#block-and-transaction-properties}}.
\newblock


\bibitem[{Ethereum Foundation}(2024b)]%
        {solidity_call}
\bibfield{author}{\bibinfo{person}{{Ethereum Foundation}}.} \bibinfo{year}{2024}\natexlab{b}.
\newblock \bibinfo{title}{Security Considerations - Sending and Receiving Ether}.
\newblock \bibinfo{howpublished}{\url{https://docs.soliditylang.org/en/latest/security-considerations.html\#sending-and-receiving-ether}}.
\newblock


\bibitem[{Ethereum Foundation}(2024c)]%
        {solidity_delegatecall}
\bibfield{author}{\bibinfo{person}{{Ethereum Foundation}}.} \bibinfo{year}{2024}\natexlab{c}.
\newblock \bibinfo{title}{Solidity by Example - Delegatecall}.
\newblock \bibinfo{howpublished}{\url{https://solidity-by-example.org/delegatecall/}}.
\newblock


\bibitem[Feist et~al\mbox{.}(2019)]%
        {feist2019slither}
\bibfield{author}{\bibinfo{person}{Josselin Feist}, \bibinfo{person}{Gustavo Grieco}, {and} \bibinfo{person}{Alex Groce}.} \bibinfo{year}{2019}\natexlab{}.
\newblock \showarticletitle{Slither: a static analysis framework for smart contracts}. In \bibinfo{booktitle}{\emph{2019 IEEE/ACM 2nd International Workshop on Emerging Trends in Software Engineering for Blockchain (WETSEB)}}. IEEE, \bibinfo{pages}{8--15}.
\newblock


\bibitem[Ferreira et~al\mbox{.}(2020)]%
        {ferreira2020smartbugs}
\bibfield{author}{\bibinfo{person}{Jo{\~a}o~F Ferreira}, \bibinfo{person}{Pedro Cruz}, \bibinfo{person}{Thomas Durieux}, {and} \bibinfo{person}{Rui Abreu}.} \bibinfo{year}{2020}\natexlab{}.
\newblock \showarticletitle{Smartbugs: A framework to analyze solidity smart contracts}. In \bibinfo{booktitle}{\emph{Proceedings of the 35th IEEE/ACM International Conference on Automated Software Engineering}}. \bibinfo{pages}{1349--1352}.
\newblock


\bibitem[Gao et~al\mbox{.}(2019b)]%
        {gao2019easyflow}
\bibfield{author}{\bibinfo{person}{Jianbo Gao}, \bibinfo{person}{Han Liu}, \bibinfo{person}{Chao Liu}, \bibinfo{person}{Qingshan Li}, \bibinfo{person}{Zhi Guan}, {and} \bibinfo{person}{Zhong Chen}.} \bibinfo{year}{2019}\natexlab{b}.
\newblock \showarticletitle{Easyflow: Keep ethereum away from overflow}. In \bibinfo{booktitle}{\emph{2019 IEEE/ACM 41st International Conference on Software Engineering: Companion Proceedings (ICSE-Companion)}}. IEEE, \bibinfo{pages}{23--26}.
\newblock


\bibitem[Gao et~al\mbox{.}(2019a)]%
        {gao2019smartembed}
\bibfield{author}{\bibinfo{person}{Zhipeng Gao}, \bibinfo{person}{Vinoj Jayasundara}, \bibinfo{person}{Lingxiao Jiang}, \bibinfo{person}{Xin Xia}, \bibinfo{person}{David Lo}, {and} \bibinfo{person}{John Grundy}.} \bibinfo{year}{2019}\natexlab{a}.
\newblock \showarticletitle{Smartembed: A tool for clone and bug detection in smart contracts through structural code embedding}. In \bibinfo{booktitle}{\emph{2019 IEEE International Conference on Software Maintenance and Evolution (ICSME)}}. IEEE, \bibinfo{pages}{394--397}.
\newblock


\bibitem[Heged{\H{u}}s(2018)]%
        {hegedHus2018towards}
\bibfield{author}{\bibinfo{person}{P{\'e}ter Heged{\H{u}}s}.} \bibinfo{year}{2018}\natexlab{}.
\newblock \showarticletitle{Towards analyzing the complexity landscape of solidity based ethereum smart contracts}. In \bibinfo{booktitle}{\emph{Proceedings of the 1st International Workshop on Emerging Trends in Software Engineering for Blockchain}}. \bibinfo{pages}{35--39}.
\newblock


\bibitem[Hewa et~al\mbox{.}(2021)]%
        {hewa2021survey}
\bibfield{author}{\bibinfo{person}{Tharaka Hewa}, \bibinfo{person}{Mika Ylianttila}, {and} \bibinfo{person}{Madhusanka Liyanage}.} \bibinfo{year}{2021}\natexlab{}.
\newblock \showarticletitle{Survey on blockchain based smart contracts: Applications, opportunities and challenges}.
\newblock \bibinfo{journal}{\emph{Journal of Network and Computer Applications}}  \bibinfo{volume}{177} (\bibinfo{year}{2021}), \bibinfo{pages}{102857}.
\newblock


\bibitem[Hu et~al\mbox{.}(2023)]%
        {hu2023large}
\bibfield{author}{\bibinfo{person}{Sihao Hu}, \bibinfo{person}{Tiansheng Huang}, \bibinfo{person}{Fatih {\.I}lhan}, \bibinfo{person}{Selim~Furkan Tekin}, {and} \bibinfo{person}{Ling Liu}.} \bibinfo{year}{2023}\natexlab{}.
\newblock \showarticletitle{Large language model-powered smart contract vulnerability detection: New perspectives}.
\newblock \bibinfo{journal}{\emph{arXiv preprint arXiv:2310.01152}} (\bibinfo{year}{2023}).
\newblock


\bibitem[Joshi et~al\mbox{.}(2015)]%
        {joshi2015likert}
\bibfield{author}{\bibinfo{person}{Ankur Joshi}, \bibinfo{person}{Saket Kale}, \bibinfo{person}{Satish Chandel}, {and} \bibinfo{person}{D~Kumar Pal}.} \bibinfo{year}{2015}\natexlab{}.
\newblock \showarticletitle{Likert scale: Explored and explained}.
\newblock \bibinfo{journal}{\emph{British journal of applied science \& technology}} \bibinfo{volume}{7}, \bibinfo{number}{4} (\bibinfo{year}{2015}), \bibinfo{pages}{396--403}.
\newblock


\bibitem[Kim et~al\mbox{.}(2023)]%
        {kim2023adaptive}
\bibfield{author}{\bibinfo{person}{Myeongsoo Kim}, \bibinfo{person}{Saurabh Sinha}, {and} \bibinfo{person}{Alessandro Orso}.} \bibinfo{year}{2023}\natexlab{}.
\newblock \showarticletitle{Adaptive rest api testing with reinforcement learning}. In \bibinfo{booktitle}{\emph{2023 38th IEEE/ACM International Conference on Automated Software Engineering (ASE)}}. IEEE, \bibinfo{pages}{446--458}.
\newblock


\bibitem[Kipf and Welling(2016)]%
        {kipf2016semi}
\bibfield{author}{\bibinfo{person}{Thomas~N Kipf} {and} \bibinfo{person}{Max Welling}.} \bibinfo{year}{2016}\natexlab{}.
\newblock \showarticletitle{Semi-supervised classification with graph convolutional networks}.
\newblock \bibinfo{journal}{\emph{arXiv preprint arXiv:1609.02907}} (\bibinfo{year}{2016}).
\newblock


\bibitem[Li et~al\mbox{.}(2024)]%
        {li2024ircoco}
\bibfield{author}{\bibinfo{person}{Bolun Li}, \bibinfo{person}{Zhihong Sun}, \bibinfo{person}{Tao Huang}, \bibinfo{person}{Hongyu Zhang}, \bibinfo{person}{Yao Wan}, \bibinfo{person}{Ge Li}, \bibinfo{person}{Zhi Jin}, {and} \bibinfo{person}{Chen Lyu}.} \bibinfo{year}{2024}\natexlab{}.
\newblock \showarticletitle{IRCoCo: Immediate Rewards-Guided Deep Reinforcement Learning for Code Completion}.
\newblock \bibinfo{journal}{\emph{Proceedings of the ACM on Software Engineering}} \bibinfo{volume}{1}, \bibinfo{number}{FSE} (\bibinfo{year}{2024}), \bibinfo{pages}{182--203}.
\newblock


\bibitem[Liu et~al\mbox{.}(2021)]%
        {liu2021smart}
\bibfield{author}{\bibinfo{person}{Zhenguang Liu}, \bibinfo{person}{Peng Qian}, \bibinfo{person}{Xiang Wang}, \bibinfo{person}{Lei Zhu}, \bibinfo{person}{Qinming He}, {and} \bibinfo{person}{Shouling Ji}.} \bibinfo{year}{2021}\natexlab{}.
\newblock \showarticletitle{Smart contract vulnerability detection: from pure neural network to interpretable graph feature and expert pattern fusion}.
\newblock \bibinfo{journal}{\emph{arXiv preprint arXiv:2106.09282}} (\bibinfo{year}{2021}).
\newblock


\bibitem[Liu et~al\mbox{.}(2023)]%
        {liu2023rethinking}
\bibfield{author}{\bibinfo{person}{Zhenguang Liu}, \bibinfo{person}{Peng Qian}, \bibinfo{person}{Jiaxu Yang}, \bibinfo{person}{Lingfeng Liu}, \bibinfo{person}{Xiaojun Xu}, \bibinfo{person}{Qinming He}, {and} \bibinfo{person}{Xiaosong Zhang}.} \bibinfo{year}{2023}\natexlab{}.
\newblock \showarticletitle{Rethinking smart contract fuzzing: Fuzzing with invocation ordering and important branch revisiting}.
\newblock \bibinfo{journal}{\emph{IEEE Transactions on Information Forensics and Security}}  \bibinfo{volume}{18} (\bibinfo{year}{2023}), \bibinfo{pages}{1237--1251}.
\newblock


\bibitem[Loshchilov and Hutter(2017)]%
        {adamw}
\bibfield{author}{\bibinfo{person}{Ilya Loshchilov} {and} \bibinfo{person}{Frank Hutter}.} \bibinfo{year}{2017}\natexlab{}.
\newblock \showarticletitle{Decoupled weight decay regularization}.
\newblock \bibinfo{journal}{\emph{arXiv preprint arXiv:1711.05101}} (\bibinfo{year}{2017}).
\newblock


\bibitem[Lu et~al\mbox{.}(2023)]%
        {lu2023llama}
\bibfield{author}{\bibinfo{person}{Junyi Lu}, \bibinfo{person}{Lei Yu}, \bibinfo{person}{Xiaojia Li}, \bibinfo{person}{Li Yang}, {and} \bibinfo{person}{Chun Zuo}.} \bibinfo{year}{2023}\natexlab{}.
\newblock \showarticletitle{Llama-reviewer: Advancing code review automation with large language models through parameter-efficient fine-tuning}. In \bibinfo{booktitle}{\emph{2023 IEEE 34th International Symposium on Software Reliability Engineering (ISSRE)}}. IEEE, \bibinfo{pages}{647--658}.
\newblock


\bibitem[Luu et~al\mbox{.}(2016)]%
        {luu2016making}
\bibfield{author}{\bibinfo{person}{Loi Luu}, \bibinfo{person}{Duc-Hiep Chu}, \bibinfo{person}{Hrishi Olickel}, \bibinfo{person}{Prateek Saxena}, {and} \bibinfo{person}{Aquinas Hobor}.} \bibinfo{year}{2016}\natexlab{}.
\newblock \showarticletitle{Making smart contracts smarter}. In \bibinfo{booktitle}{\emph{Proceedings of the 2016 ACM SIGSAC conference on computer and communications security}}. \bibinfo{pages}{254--269}.
\newblock


\bibitem[Ma et~al\mbox{.}(2024)]%
        {ma2024combining}
\bibfield{author}{\bibinfo{person}{Wei Ma}, \bibinfo{person}{Daoyuan Wu}, \bibinfo{person}{Yuqiang Sun}, \bibinfo{person}{Tianwen Wang}, \bibinfo{person}{Shangqing Liu}, \bibinfo{person}{Jian Zhang}, \bibinfo{person}{Yue Xue}, {and} \bibinfo{person}{Yang Liu}.} \bibinfo{year}{2024}\natexlab{}.
\newblock \showarticletitle{Combining Fine-Tuning and LLM-based Agents for Intuitive Smart Contract Auditing with Justifications}.
\newblock \bibinfo{journal}{\emph{arXiv preprint arXiv:2403.16073}} (\bibinfo{year}{2024}).
\newblock


\bibitem[Mehar et~al\mbox{.}(2019)]%
        {mehar2019understanding}
\bibfield{author}{\bibinfo{person}{Muhammad~Izhar Mehar}, \bibinfo{person}{Charles~Louis Shier}, \bibinfo{person}{Alana Giambattista}, \bibinfo{person}{Elgar Gong}, \bibinfo{person}{Gabrielle Fletcher}, \bibinfo{person}{Ryan Sanayhie}, \bibinfo{person}{Henry~M Kim}, {and} \bibinfo{person}{Marek Laskowski}.} \bibinfo{year}{2019}\natexlab{}.
\newblock \showarticletitle{Understanding a revolutionary and flawed grand experiment in blockchain: the DAO attack}.
\newblock \bibinfo{journal}{\emph{Journal of Cases on Information Technology (JCIT)}} \bibinfo{volume}{21}, \bibinfo{number}{1} (\bibinfo{year}{2019}), \bibinfo{pages}{19--32}.
\newblock


\bibitem[Mistral(2024)]%
        {mistrallarge2}
\bibfield{author}{\bibinfo{person}{Mistral}.} \bibinfo{year}{2024}\natexlab{}.
\newblock \bibinfo{title}{Large Enough | Mistral AI | Frontier AI in your hands}.
\newblock
\urldef\tempurl%
\url{https://mistral.ai/news/mistral-large-2407/}
\showURL{%
\tempurl}


\bibitem[Mueller(2017)]%
        {mueller2017mythril}
\bibfield{author}{\bibinfo{person}{B Mueller}.} \bibinfo{year}{2017}\natexlab{}.
\newblock \bibinfo{title}{Mythril-Reversing and bug hunting framework for the Ethereum blockchain}.
\newblock


\bibitem[Nashaat and Miller(2024)]%
        {nashaat2024towards}
\bibfield{author}{\bibinfo{person}{Mona Nashaat} {and} \bibinfo{person}{James Miller}.} \bibinfo{year}{2024}\natexlab{}.
\newblock \showarticletitle{Towards Efficient Fine-tuning of Language Models with Organizational Data for Automated Software Review}.
\newblock \bibinfo{journal}{\emph{IEEE Transactions on Software Engineering}} (\bibinfo{year}{2024}).
\newblock


\bibitem[Nguyen et~al\mbox{.}(2020)]%
        {nguyen2020sfuzz}
\bibfield{author}{\bibinfo{person}{Tai~D Nguyen}, \bibinfo{person}{Long~H Pham}, \bibinfo{person}{Jun Sun}, \bibinfo{person}{Yun Lin}, {and} \bibinfo{person}{Quang~Tran Minh}.} \bibinfo{year}{2020}\natexlab{}.
\newblock \showarticletitle{sfuzz: An efficient adaptive fuzzer for solidity smart contracts}. In \bibinfo{booktitle}{\emph{Proceedings of the ACM/IEEE 42nd International Conference on Software Engineering}}. \bibinfo{pages}{778--788}.
\newblock


\bibitem[Nikanjam et~al\mbox{.}(2022)]%
        {nikanjam2022faults}
\bibfield{author}{\bibinfo{person}{Amin Nikanjam}, \bibinfo{person}{Mohammad~Mehdi Morovati}, \bibinfo{person}{Foutse Khomh}, {and} \bibinfo{person}{Houssem Ben~Braiek}.} \bibinfo{year}{2022}\natexlab{}.
\newblock \showarticletitle{Faults in deep reinforcement learning programs: a taxonomy and a detection approach}.
\newblock \bibinfo{journal}{\emph{Automated software engineering}} \bibinfo{volume}{29}, \bibinfo{number}{1} (\bibinfo{year}{2022}), \bibinfo{pages}{8}.
\newblock


\bibitem[OpenAI(2024)]%
        {openai2024gpt4turbo}
\bibfield{author}{\bibinfo{person}{OpenAI}.} \bibinfo{year}{2024}\natexlab{}.
\newblock \bibinfo{title}{GPT-4-Turbo}.
\newblock \bibinfo{howpublished}{\url{https://platform.openai.com/docs/models/gpt-4-and-gpt-4-turbo}}.
\newblock


\bibitem[{OWASP Foundation}(2023)]%
        {owasp2023}
\bibfield{author}{\bibinfo{person}{{OWASP Foundation}}.} \bibinfo{year}{2023}\natexlab{}.
\newblock \bibinfo{title}{{OWASP Smart Contract Top 10}}.
\newblock \bibinfo{howpublished}{\url{https://owasp.org/www-project-smart-contract-top-10/}}.
\newblock
\newblock
\shownote{Accessed: 2024-12-04}.


\bibitem[Praitheeshan et~al\mbox{.}(2019)]%
        {praitheeshan2019security}
\bibfield{author}{\bibinfo{person}{Purathani Praitheeshan}, \bibinfo{person}{Lei Pan}, \bibinfo{person}{Jiangshan Yu}, \bibinfo{person}{Joseph Liu}, {and} \bibinfo{person}{Robin Doss}.} \bibinfo{year}{2019}\natexlab{}.
\newblock \showarticletitle{Security analysis methods on ethereum smart contract vulnerabilities: a survey}.
\newblock \bibinfo{journal}{\emph{arXiv preprint arXiv:1908.08605}} (\bibinfo{year}{2019}).
\newblock


\bibitem[{Protofire}(2020)]%
        {protofire2020solhint}
\bibfield{author}{\bibinfo{person}{{Protofire}}.} \bibinfo{year}{2020}\natexlab{}.
\newblock \bibinfo{title}{Solhint: A linting utility for {Solidity} code}.
\newblock \bibinfo{howpublished}{\url{https://github.com/protofire/solhint}}.
\newblock
\newblock
\shownote{GitHub repository with over 1000 stars, Used by over 54,000 repositories}.


\bibitem[Qian et~al\mbox{.}(2020)]%
        {qian2020towards}
\bibfield{author}{\bibinfo{person}{Peng Qian}, \bibinfo{person}{Zhenguang Liu}, \bibinfo{person}{Qinming He}, \bibinfo{person}{Roger Zimmermann}, {and} \bibinfo{person}{Xun Wang}.} \bibinfo{year}{2020}\natexlab{}.
\newblock \showarticletitle{Towards automated reentrancy detection for smart contracts based on sequential models}.
\newblock \bibinfo{journal}{\emph{IEEE Access}}  \bibinfo{volume}{8} (\bibinfo{year}{2020}), \bibinfo{pages}{19685--19695}.
\newblock


\bibitem[Qian et~al\mbox{.}(2023)]%
        {qian2023cross}
\bibfield{author}{\bibinfo{person}{Peng Qian}, \bibinfo{person}{Zhenguang Liu}, \bibinfo{person}{Yifang Yin}, {and} \bibinfo{person}{Qinming He}.} \bibinfo{year}{2023}\natexlab{}.
\newblock \showarticletitle{Cross-modality mutual learning for enhancing smart contract vulnerability detection on bytecode}. In \bibinfo{booktitle}{\emph{Proceedings of the ACM Web Conference 2023}}. \bibinfo{pages}{2220--2229}.
\newblock


\bibitem[Qwen(2024)]%
        {qwen2.5}
\bibfield{author}{\bibinfo{person}{Qwen}.} \bibinfo{year}{2024}\natexlab{}.
\newblock \bibinfo{title}{Qwen2.5: A Party of Foundation Models}.
\newblock
\urldef\tempurl%
\url{https://qwenlm.github.io/blog/qwen2.5/}
\showURL{%
\tempurl}


\bibitem[Rafailov et~al\mbox{.}(2023)]%
        {rafailov2023direct}
\bibfield{author}{\bibinfo{person}{Rafael Rafailov}, \bibinfo{person}{Archit Sharma}, \bibinfo{person}{Eric Mitchell}, \bibinfo{person}{Christopher~D Manning}, \bibinfo{person}{Stefano Ermon}, {and} \bibinfo{person}{Chelsea Finn}.} \bibinfo{year}{2023}\natexlab{}.
\newblock \showarticletitle{Direct preference optimization: Your language model is secretly a reward model}.
\newblock \bibinfo{journal}{\emph{Advances in Neural Information Processing Systems}}  \bibinfo{volume}{36} (\bibinfo{year}{2023}), \bibinfo{pages}{53728--53741}.
\newblock


\bibitem[Rasley et~al\mbox{.}(2020)]%
        {rasley2020deepspeed}
\bibfield{author}{\bibinfo{person}{Jeff Rasley}, \bibinfo{person}{Samyam Rajbhandari}, \bibinfo{person}{Olatunji Ruwase}, {and} \bibinfo{person}{Yuxiong He}.} \bibinfo{year}{2020}\natexlab{}.
\newblock \showarticletitle{Deepspeed: System optimizations enable training deep learning models with over 100 billion parameters}. In \bibinfo{booktitle}{\emph{Proceedings of the 26th ACM SIGKDD International Conference on Knowledge Discovery \& Data Mining}}. \bibinfo{pages}{3505--3506}.
\newblock


\bibitem[Reddy et~al\mbox{.}(2020)]%
        {reddy2020quickly}
\bibfield{author}{\bibinfo{person}{Sameer Reddy}, \bibinfo{person}{Caroline Lemieux}, \bibinfo{person}{Rohan Padhye}, {and} \bibinfo{person}{Koushik Sen}.} \bibinfo{year}{2020}\natexlab{}.
\newblock \showarticletitle{Quickly generating diverse valid test inputs with reinforcement learning}. In \bibinfo{booktitle}{\emph{Proceedings of the ACM/IEEE 42nd International Conference on Software Engineering}}. \bibinfo{pages}{1410--1421}.
\newblock


\bibitem[Romdhana et~al\mbox{.}(2022)]%
        {romdhana2022deep}
\bibfield{author}{\bibinfo{person}{Andrea Romdhana}, \bibinfo{person}{Alessio Merlo}, \bibinfo{person}{Mariano Ceccato}, {and} \bibinfo{person}{Paolo Tonella}.} \bibinfo{year}{2022}\natexlab{}.
\newblock \showarticletitle{Deep reinforcement learning for black-box testing of android apps}.
\newblock \bibinfo{journal}{\emph{ACM Transactions on Software Engineering and Methodology (TOSEM)}} \bibinfo{volume}{31}, \bibinfo{number}{4} (\bibinfo{year}{2022}), \bibinfo{pages}{1--29}.
\newblock


\bibitem[Storhaug et~al\mbox{.}(2023)]%
        {storhaug2023efficient}
\bibfield{author}{\bibinfo{person}{Andr{\'e} Storhaug}, \bibinfo{person}{Jingyue Li}, {and} \bibinfo{person}{Tianyuan Hu}.} \bibinfo{year}{2023}\natexlab{}.
\newblock \showarticletitle{Efficient avoidance of vulnerabilities in auto-completed smart contract code using vulnerability-constrained decoding}. In \bibinfo{booktitle}{\emph{2023 IEEE 34th International Symposium on Software Reliability Engineering (ISSRE)}}. IEEE, \bibinfo{pages}{683--693}.
\newblock


\bibitem[Sun et~al\mbox{.}(2024a)]%
        {sun2024llm4vuln}
\bibfield{author}{\bibinfo{person}{Yuqiang Sun}, \bibinfo{person}{Daoyuan Wu}, \bibinfo{person}{Yue Xue}, \bibinfo{person}{Han Liu}, \bibinfo{person}{Wei Ma}, \bibinfo{person}{Lyuye Zhang}, \bibinfo{person}{Yang Liu}, {and} \bibinfo{person}{Yingjiu Li}.} \bibinfo{year}{2024}\natexlab{a}.
\newblock \showarticletitle{Llm4vuln: A unified evaluation framework for decoupling and enhancing llms' vulnerability reasoning}.
\newblock \bibinfo{journal}{\emph{arXiv preprint arXiv:2401.16185}} (\bibinfo{year}{2024}).
\newblock


\bibitem[Sun et~al\mbox{.}(2024b)]%
        {sun2024gptscan}
\bibfield{author}{\bibinfo{person}{Yuqiang Sun}, \bibinfo{person}{Daoyuan Wu}, \bibinfo{person}{Yue Xue}, \bibinfo{person}{Han Liu}, \bibinfo{person}{Haijun Wang}, \bibinfo{person}{Zhengzi Xu}, \bibinfo{person}{Xiaofei Xie}, {and} \bibinfo{person}{Yang Liu}.} \bibinfo{year}{2024}\natexlab{b}.
\newblock \showarticletitle{GPTScan: Detecting Logic Vulnerabilities in Smart Contracts by Combining GPT with Program Analysis}.
\newblock \bibinfo{journal}{\emph{Proc. IEEE/ACM ICSE}} (\bibinfo{year}{2024}).
\newblock


\bibitem[Swan(2015)]%
        {swan2015blockchain}
\bibfield{author}{\bibinfo{person}{Melanie Swan}.} \bibinfo{year}{2015}\natexlab{}.
\newblock \bibinfo{booktitle}{\emph{Blockchain: Blueprint for a new economy}}.
\newblock \bibinfo{publisher}{" O'Reilly Media, Inc."}.
\newblock


\bibitem[Tikhomirov et~al\mbox{.}(2018)]%
        {tikhomirov2018smartcheck}
\bibfield{author}{\bibinfo{person}{Sergei Tikhomirov}, \bibinfo{person}{Ekaterina Voskresenskaya}, \bibinfo{person}{Ivan Ivanitskiy}, \bibinfo{person}{Ramil Takhaviev}, \bibinfo{person}{Evgeny Marchenko}, {and} \bibinfo{person}{Yaroslav Alexandrov}.} \bibinfo{year}{2018}\natexlab{}.
\newblock \showarticletitle{Smartcheck: Static analysis of ethereum smart contracts}. In \bibinfo{booktitle}{\emph{Proceedings of the 1st International Workshop on Emerging Trends in Software Engineering for Blockchain}}. \bibinfo{pages}{9--16}.
\newblock


\bibitem[Torres et~al\mbox{.}(2021)]%
        {torres2021confuzzius}
\bibfield{author}{\bibinfo{person}{Christof~Ferreira Torres}, \bibinfo{person}{Antonio~Ken Iannillo}, \bibinfo{person}{Arthur Gervais}, {and} \bibinfo{person}{Radu State}.} \bibinfo{year}{2021}\natexlab{}.
\newblock \showarticletitle{Confuzzius: A data dependency-aware hybrid fuzzer for smart contracts}. In \bibinfo{booktitle}{\emph{2021 IEEE European Symposium on Security and Privacy (EuroS\&P)}}. IEEE, \bibinfo{pages}{103--119}.
\newblock


\bibitem[Torres et~al\mbox{.}(2018)]%
        {torres2018osiris}
\bibfield{author}{\bibinfo{person}{Christof~Ferreira Torres}, \bibinfo{person}{Julian Sch{\"u}tte}, {and} \bibinfo{person}{Radu State}.} \bibinfo{year}{2018}\natexlab{}.
\newblock \showarticletitle{Osiris: Hunting for integer bugs in ethereum smart contracts}. In \bibinfo{booktitle}{\emph{Proceedings of the 34th Annual Computer Security Applications Conference}}. \bibinfo{pages}{664--676}.
\newblock


\bibitem[Tsankov et~al\mbox{.}(2018)]%
        {tsankov2018securify}
\bibfield{author}{\bibinfo{person}{Petar Tsankov}, \bibinfo{person}{Andrei Dan}, \bibinfo{person}{Dana Drachsler-Cohen}, \bibinfo{person}{Arthur Gervais}, \bibinfo{person}{Florian Buenzli}, {and} \bibinfo{person}{Martin Vechev}.} \bibinfo{year}{2018}\natexlab{}.
\newblock \showarticletitle{Securify: Practical security analysis of smart contracts}. In \bibinfo{booktitle}{\emph{Proceedings of the 2018 ACM SIGSAC Conference on Computer and Communications Security}}. \bibinfo{pages}{67--82}.
\newblock


\bibitem[Veloso(2021)]%
        {veloso2021conkas}
\bibfield{author}{\bibinfo{person}{Nuno Veloso}.} \bibinfo{year}{2021}\natexlab{}.
\newblock \bibinfo{title}{Conkas}.
\newblock \bibinfo{howpublished}{\url{https://github.com/nveloso/conkas}}.
\newblock
\newblock
\shownote{GitHub repository}.


\bibitem[Wang et~al\mbox{.}(2023)]%
        {wang2023generating}
\bibfield{author}{\bibinfo{person}{Chong Wang}, \bibinfo{person}{Yiling Lou}, \bibinfo{person}{Junwei Liu}, {and} \bibinfo{person}{Xin Peng}.} \bibinfo{year}{2023}\natexlab{}.
\newblock \showarticletitle{Generating variable explanations via zero-shot prompt learning}. In \bibinfo{booktitle}{\emph{2023 38th IEEE/ACM International Conference on Automated Software Engineering (ASE)}}. IEEE, \bibinfo{pages}{748--760}.
\newblock


\bibitem[Wang et~al\mbox{.}(2018)]%
        {wang2018ccaligner}
\bibfield{author}{\bibinfo{person}{Pengcheng Wang}, \bibinfo{person}{Jeffrey Svajlenko}, \bibinfo{person}{Yanzhao Wu}, \bibinfo{person}{Yun Xu}, {and} \bibinfo{person}{Chanchal~K Roy}.} \bibinfo{year}{2018}\natexlab{}.
\newblock \showarticletitle{CCAligner: a token based large-gap clone detector}. In \bibinfo{booktitle}{\emph{Proceedings of the 40th International Conference on Software Engineering}}. \bibinfo{pages}{1066--1077}.
\newblock


\bibitem[Wang et~al\mbox{.}(2022)]%
        {wang2022self}
\bibfield{author}{\bibinfo{person}{Xuezhi Wang}, \bibinfo{person}{Jason Wei}, \bibinfo{person}{Dale Schuurmans}, \bibinfo{person}{Quoc Le}, \bibinfo{person}{Ed Chi}, \bibinfo{person}{Sharan Narang}, \bibinfo{person}{Aakanksha Chowdhery}, {and} \bibinfo{person}{Denny Zhou}.} \bibinfo{year}{2022}\natexlab{}.
\newblock \showarticletitle{Self-consistency improves chain of thought reasoning in language models}.
\newblock \bibinfo{journal}{\emph{arXiv preprint arXiv:2203.11171}} (\bibinfo{year}{2022}).
\newblock


\bibitem[Wang et~al\mbox{.}(2021)]%
        {wang2021codet5}
\bibfield{author}{\bibinfo{person}{Yue Wang}, \bibinfo{person}{Weishi Wang}, \bibinfo{person}{Shafiq Joty}, {and} \bibinfo{person}{Steven~CH Hoi}.} \bibinfo{year}{2021}\natexlab{}.
\newblock \showarticletitle{CodeT5: Identifier-aware Unified Pre-trained Encoder-Decoder Models for Code Understanding and Generation}. In \bibinfo{booktitle}{\emph{Proceedings of the 2021 Conference on Empirical Methods in Natural Language Processing}}. \bibinfo{pages}{8696--8708}.
\newblock


\bibitem[Wang et~al\mbox{.}(2024)]%
        {wang2024rlcoder}
\bibfield{author}{\bibinfo{person}{Yanlin Wang}, \bibinfo{person}{Yanli Wang}, \bibinfo{person}{Daya Guo}, \bibinfo{person}{Jiachi Chen}, \bibinfo{person}{Ruikai Zhang}, \bibinfo{person}{Yuchi Ma}, {and} \bibinfo{person}{Zibin Zheng}.} \bibinfo{year}{2024}\natexlab{}.
\newblock \showarticletitle{Rlcoder: Reinforcement learning for repository-level code completion}.
\newblock \bibinfo{journal}{\emph{arXiv preprint arXiv:2407.19487}} (\bibinfo{year}{2024}).
\newblock


\bibitem[Wei et~al\mbox{.}(2024)]%
        {wei2024leveraging}
\bibfield{author}{\bibinfo{person}{Zhiyuan Wei}, \bibinfo{person}{Jing Sun}, \bibinfo{person}{Zijian Zhang}, \bibinfo{person}{Xianhao Zhang}, {and} \bibinfo{person}{Meng Li}.} \bibinfo{year}{2024}\natexlab{}.
\newblock \showarticletitle{Leveraging Fine-Tuned Language Models for Efficient and Accurate Smart Contract Auditing}.
\newblock \bibinfo{journal}{\emph{arXiv preprint arXiv:2410.13918}} (\bibinfo{year}{2024}).
\newblock


\bibitem[Wood et~al\mbox{.}(2014)]%
        {wood2014ethereum}
\bibfield{author}{\bibinfo{person}{Gavin Wood} {et~al\mbox{.}}} \bibinfo{year}{2014}\natexlab{}.
\newblock \showarticletitle{Ethereum: A secure decentralised generalised transaction ledger}.
\newblock \bibinfo{journal}{\emph{Ethereum project yellow paper}} \bibinfo{volume}{151}, \bibinfo{number}{2014} (\bibinfo{year}{2014}), \bibinfo{pages}{1--32}.
\newblock


\bibitem[Wu et~al\mbox{.}(2021)]%
        {wu2021peculiar}
\bibfield{author}{\bibinfo{person}{Hongjun Wu}, \bibinfo{person}{Zhuo Zhang}, \bibinfo{person}{Shangwen Wang}, \bibinfo{person}{Yan Lei}, \bibinfo{person}{Bo Lin}, \bibinfo{person}{Yihao Qin}, \bibinfo{person}{Haoyu Zhang}, {and} \bibinfo{person}{Xiaoguang Mao}.} \bibinfo{year}{2021}\natexlab{}.
\newblock \showarticletitle{Peculiar: Smart contract vulnerability detection based on crucial data flow graph and pre-training techniques}. In \bibinfo{booktitle}{\emph{2021 IEEE 32nd International Symposium on Software Reliability Engineering (ISSRE)}}. IEEE, \bibinfo{pages}{378--389}.
\newblock


\bibitem[Yang et~al\mbox{.}(2024)]%
        {yang2024qwen2}
\bibfield{author}{\bibinfo{person}{An Yang}, \bibinfo{person}{Baosong Yang}, \bibinfo{person}{Binyuan Hui}, \bibinfo{person}{Bo Zheng}, \bibinfo{person}{Bowen Yu}, \bibinfo{person}{Chang Zhou}, \bibinfo{person}{Chengpeng Li}, \bibinfo{person}{Chengyuan Li}, \bibinfo{person}{Dayiheng Liu}, \bibinfo{person}{Fei Huang}, {et~al\mbox{.}}} \bibinfo{year}{2024}\natexlab{}.
\newblock \showarticletitle{Qwen2 technical report}.
\newblock \bibinfo{journal}{\emph{arXiv preprint arXiv:2407.10671}} (\bibinfo{year}{2024}).
\newblock


\bibitem[Yu et~al\mbox{.}(2024)]%
        {yu2024smart}
\bibfield{author}{\bibinfo{person}{Lei Yu}, \bibinfo{person}{Shiqi Chen}, \bibinfo{person}{Hang Yuan}, \bibinfo{person}{Peng Wang}, \bibinfo{person}{Zhirong Huang}, \bibinfo{person}{Jingyuan Zhang}, \bibinfo{person}{Chenjie Shen}, \bibinfo{person}{Fengjun Zhang}, \bibinfo{person}{Li Yang}, {and} \bibinfo{person}{Jiajia Ma}.} \bibinfo{year}{2024}\natexlab{}.
\newblock \showarticletitle{Smart-LLaMA: Two-Stage Post-Training of Large Language Models for Smart Contract Vulnerability Detection and Explanation}.
\newblock \bibinfo{journal}{\emph{arXiv preprint arXiv:2411.06221}} (\bibinfo{year}{2024}).
\newblock


\bibitem[Yu et~al\mbox{.}(2023a)]%
        {yu2023pscvfinder}
\bibfield{author}{\bibinfo{person}{Lei Yu}, \bibinfo{person}{Junyi Lu}, \bibinfo{person}{Xianglong Liu}, \bibinfo{person}{Li Yang}, \bibinfo{person}{Fengjun Zhang}, {and} \bibinfo{person}{Jiajia Ma}.} \bibinfo{year}{2023}\natexlab{a}.
\newblock \showarticletitle{PSCVFinder: A Prompt-Tuning Based Framework for Smart Contract Vulnerability Detection}. In \bibinfo{booktitle}{\emph{2023 IEEE 34th International Symposium on Software Reliability Engineering (ISSRE)}}. IEEE, \bibinfo{pages}{556--567}.
\newblock


\bibitem[Yu et~al\mbox{.}(2023b)]%
        {yu2023money}
\bibfield{author}{\bibinfo{person}{Lei Yu}, \bibinfo{person}{Fengjun Zhang}, \bibinfo{person}{Jiajia Ma}, \bibinfo{person}{Li Yang}, \bibinfo{person}{Yuanzhe Yang}, {and} \bibinfo{person}{Wei Jia}.} \bibinfo{year}{2023}\natexlab{b}.
\newblock \showarticletitle{Who Are the Money Launderers? Money Laundering Detection on Blockchain via Mutual Learning-Based Graph Neural Network}. In \bibinfo{booktitle}{\emph{2023 International Joint Conference on Neural Networks (IJCNN)}}. IEEE, \bibinfo{pages}{1--8}.
\newblock


\bibitem[Zhang et~al\mbox{.}(2023b)]%
        {zhang2023critical}
\bibfield{author}{\bibinfo{person}{Quanjun Zhang}, \bibinfo{person}{Tongke Zhang}, \bibinfo{person}{Juan Zhai}, \bibinfo{person}{Chunrong Fang}, \bibinfo{person}{Bowen Yu}, \bibinfo{person}{Weisong Sun}, {and} \bibinfo{person}{Zhenyu Chen}.} \bibinfo{year}{2023}\natexlab{b}.
\newblock \showarticletitle{A critical review of large language model on software engineering: An example from chatgpt and automated program repair}.
\newblock \bibinfo{journal}{\emph{arXiv preprint arXiv:2310.08879}} (\bibinfo{year}{2023}).
\newblock


\bibitem[Zhang et~al\mbox{.}(2023a)]%
        {zhang2023demystifying}
\bibfield{author}{\bibinfo{person}{Zhuo Zhang}, \bibinfo{person}{Brian Zhang}, \bibinfo{person}{Wen Xu}, {and} \bibinfo{person}{Zhiqiang Lin}.} \bibinfo{year}{2023}\natexlab{a}.
\newblock \showarticletitle{Demystifying exploitable bugs in smart contracts}. In \bibinfo{booktitle}{\emph{2023 IEEE/ACM 45th International Conference on Software Engineering (ICSE)}}. IEEE, \bibinfo{pages}{615--627}.
\newblock


\bibitem[Zheng et~al\mbox{.}(2024)]%
        {zheng2024llamafactory}
\bibfield{author}{\bibinfo{person}{Yaowei Zheng}, \bibinfo{person}{Richong Zhang}, \bibinfo{person}{Junhao Zhang}, \bibinfo{person}{Yanhan Ye}, {and} \bibinfo{person}{Zheyan Luo}.} \bibinfo{year}{2024}\natexlab{}.
\newblock \showarticletitle{Llamafactory: Unified efficient fine-tuning of 100+ language models}.
\newblock \bibinfo{journal}{\emph{arXiv preprint arXiv:2403.13372}} (\bibinfo{year}{2024}).
\newblock


\bibitem[Zhuang et~al\mbox{.}(2020)]%
        {zhuang2020smart}
\bibfield{author}{\bibinfo{person}{Yuan Zhuang}, \bibinfo{person}{Zhenguang Liu}, \bibinfo{person}{Peng Qian}, \bibinfo{person}{Qi Liu}, \bibinfo{person}{Xiang Wang}, {and} \bibinfo{person}{Qinming He}.} \bibinfo{year}{2020}\natexlab{}.
\newblock \showarticletitle{Smart Contract Vulnerability Detection using Graph Neural Network.}. In \bibinfo{booktitle}{\emph{IJCAI}}. \bibinfo{pages}{3283--3290}.
\newblock


\bibitem[Zou et~al\mbox{.}(2019)]%
        {zou2019smart}
\bibfield{author}{\bibinfo{person}{Weiqin Zou}, \bibinfo{person}{David Lo}, \bibinfo{person}{Pavneet~Singh Kochhar}, \bibinfo{person}{Xuan-Bach~Dinh Le}, \bibinfo{person}{Xin Xia}, \bibinfo{person}{Yang Feng}, \bibinfo{person}{Zhenyu Chen}, {and} \bibinfo{person}{Baowen Xu}.} \bibinfo{year}{2019}\natexlab{}.
\newblock \showarticletitle{Smart contract development: Challenges and opportunities}.
\newblock \bibinfo{journal}{\emph{IEEE Transactions on Software Engineering}} \bibinfo{volume}{47}, \bibinfo{number}{10} (\bibinfo{year}{2019}), \bibinfo{pages}{2084--2106}.
\newblock


\end{thebibliography}
\end{document}